\documentclass[aps,prb,twocolumn,superscriptaddress]{revtex4-2}
\usepackage{bm}                     
\usepackage{graphicx}               
\usepackage{graphics}               
\usepackage{amsmath}                
\usepackage{amsfonts}               
\usepackage{amssymb}                
\usepackage{makeidx}                
\usepackage{wasysym}                
\usepackage{bbold}                  
\usepackage[makeroom]{cancel}       
\usepackage[utf8]{inputenc}         
\usepackage[normalem]{ulem}         
\usepackage{xcolor}                 
\usepackage{soul}                   
\usepackage{float}                  
\usepackage{physics}                
\usepackage{IEEEtrantools}          
\usepackage{mathrsfs} 

\usepackage[T1]{fontenc}
\usepackage[utf8]{inputenc}

\newcommand{\be}{\begin{equation}}
\newcommand{\ee}{\end{equation}}
\newcommand{\ba}   {\begin{eqnarray}}
\newcommand{\ea}   {\end{eqnarray}}

\usepackage[unicode=true,pdfusetitle,
 bookmarks=true,bookmarksnumbered=false,bookmarksopen=false,
 breaklinks=false,pdfborder={0 0 1},backref=false,colorlinks=false]
 {hyperref}
\hypersetup{
 colorlinks,linkcolor=blue,citecolor=blue,urlcolor=blue}

\definecolor{navy}{RGB}{0,0,128}
\definecolor{royalblue}{RGB}{65,105,225}
\definecolor{darkorange}{RGB}{204,85,0} 
\definecolor{forestgreen}{RGB}{34,139,34}

\begingroup

\footnotetext{These authors contributed equally to this work.}
\endgroup

\begin{document}

\preprint{APS/123-QED}
\title{Staggered orbital magnetization from itinerant electrons: orbital antiferro- and ferrimagnetic phases}

\author{Lucas L. Lage $^{\P}$}
\affiliation{Instituto de F\'\i sica, Universidade Federal Fluminense, 24210-346 Niter\'oi RJ, Brazil} 

\author{Kevin J. U. Vidarte $^{\P}$}
\email{kevin.urcia@inl.int}
\affiliation{International Iberian Nanotechnology Laboratory (INL), Av. Mestre José Veiga, 4715-330 Braga, Portugal}

\author{Tarik P. Cysne}
\email{tarik.cysne@gmail.com}
\affiliation{Instituto de F\'\i sica, Universidade Federal Fluminense, 24210-346 Niter\'oi RJ, Brazil} 

\author{Tatiana G. Rappoport}
\email{tatiana.rappoport@inl.int}
\affiliation{International Iberian Nanotechnology Laboratory (INL), Av. Mestre José Veiga, 4715-330 Braga, Portugal}
\affiliation{Centro Brasileiro de Pesquisas Físicas (CBPF), Rua Dr Xavier Sigaud 150, Urca, 22290-180, Rio de Janeiro-RJ, Brazil}
\affiliation{Physics Center of Minho and Porto Universities (CF-UM-UP),Campus of Gualtar, 4710-057, Braga, Portugal}

\author{A. Latg\'e}
\affiliation{Instituto de F\'\i sica, Universidade Federal Fluminense, 24210-346 Niter\'oi RJ, Brazil} 

\author{R. B. Muniz}
\affiliation{Instituto de F\'\i sica, Universidade Federal Fluminense, 24210-346 Niter\'oi RJ, Brazil}

\date{\today}

\begin{abstract}
Because electronic orbital angular momentum in solids is inherently non-local, its contribution to magnetism is usually cast in terms of a net orbital magnetization. Here, we show that itinerant electrons can generate orbital magnetic phases with ferromagnetic, antiferromagnetic, or ferrimagnetic orders. We demonstrate this possibility in a honeycomb lattice, using both the standard and a modified Haldane model. Employing real-space formulations, we decompose the itinerant orbital magnetization into sublattice contributions, $M_A$ and $M_B$. Their net ($M_z=M_A+M_B$) and staggered ($M_z^s=M_A-M_B$) combinations are then used to identify the orbital order. By varying the sublattice potential and the Fermi energy, we find distinct regimes: a (PT)-symmetric orbital antiferromagnet in the modified Haldane model, an orbital ferromagnet in the standard Haldane model, ferrimagnetic metallic states where net and staggered orbital magnetizations coexist, and insulating regimes in which the ferro- and antiferromagnetic orbital characters can be interchanged.These findings are explained by a low-energy theory in terms of two distinct valley mechanisms: valley-dependent Dirac masses in the standard Haldane model and valley-dependent energy shifts in its modified version.
\end{abstract}


\maketitle

 

{\bf Introduction:}
Orbital angular momentum has become a central ingredient in the description of magnetic and transport phenomena in solids. In orbitronics \cite{Go2021,Jo2024,Atencia2024,Cysne2025,Ando2025,Fukami2025,Wang2024}, it can be generated, transported, and converted into spin or magnetic torques through effects such as the orbital Hall effect \cite{Choi2023, Bernevig2005, Cysne2021,Go2018,Abrao2025, Mele2019, Bhowal2021}, orbital Edelstein effect \cite{Johansson2024, Salemi2019, Nikolaev2024, Xiao2026}, and orbital torque \cite{Lee2021,Yang2024, Go2020, Go2023, Santos2023}. Recent experiments have further shown that orbital angular momentum can play an active role in magnetic materials, including antiferromagnetic systems such as CoO, where orbital torques and orbital magnetoresistance have been reported \cite{Ding2025CoO, Schmitt2026CoO}. These developments indicate that orbital angular momentum is not merely a secondary correction to spin magnetism, but an electronic degree of freedom capable of producing measurable magnetic responses.

However, most discussions of orbital magnetism focus either on the total orbital magnetization or on orbital currents. Much less is known about whether itinerant orbital magnetization can organize into compensated or partially compensated magnetic patterns analogous to spin antiferromagnets and ferrimagnets. This question has gained renewed relevance with the broader search for unconventional magnetic order. In spin systems, altermagnetism has shown that compensated magnetic order can display momentum-dependent splittings and transport signatures beyond those of conventional antiferromagnets \cite{Smejkal2020,Hayami2019,Yuan2020,Smejkal2022}. In the orbital sector, recent works have proposed non-ferromagnetic forms of orbital order, including $p$-wave orbital magnetism and orbital altermagnetism \cite{Li2026,Pan2026}. Our focus here is different: we ask whether itinerant orbital magnetization itself can display ferro-, antiferro-, and ferrimagnetic character when resolved over the sublattices of a crystal.

This question is subtle because orbital magnetization in an itinerant solid is not generally a sum of localized atomic orbital moments. In an atom-centered picture, one may assign orbital angular momentum to each site and build ferro- or antiferro-orbital patterns by analogy with localized spins. In a crystalline solid, however, orbital magnetization contains contributions from the circulation of Bloch electrons over the whole system. Its proper description is given by the modern theory of orbital magnetization, in which Berry-phase and Fermi-energy contributions are essential \cite{Xiao2005,Thonhauser2005,Ceresoli2006,Souza2008}. Therefore, identifying orbital antiferromagnetism or ferrimagnetism in the itinerant regime requires a sublattice-resolved quantity that remains defined from the itinerant orbital magnetization, rather than from local atomic moments.

The standard \cite{Haldane1988} and modified \cite{Colomes2018} Haldane models provide a minimal setting to address this question. They are defined on the same honeycomb lattice and differ only in the convention adopted for the complex next-nearest-neighbor hopping phase. Nevertheless, this apparently small change has a qualitative consequence in the low-energy theory. In the standard Haldane model, the complex hopping acts as a valley-dependent Dirac mass, whereas in the modified Haldane model it acts as a valley-dependent energy shift. The two models, therefore, offer a controlled way of comparing distinct itinerant mechanisms for generating orbital magnetization while keeping the lattice and sublattice structure fixed.

Our analysis relies on real-space formulations of orbital magnetization. We compute sublattice-resolved contributions $M_A$ and $M_B$ using both the local-marker approach \cite{Bianco2013} and the Chebyshev spectral method \cite{Vidarte2026}. Their sum and difference,
$M_z=M_A+M_B$ and $M_z^s=M_A-M_B$, provide the quantities used to identify net and staggered orbital magnetic responses.

With this approach, the same pair of Haldane-type models can be used to access different orbital magnetic regimes. A finite $M_z$ with vanishing $M_z^s$ corresponds to an itinerant orbital ferromagnetic state (OFM), a finite $M_z^s$ with vanishing $M_z$ corresponds to an itinerant orbital antiferromagnetic state (OAFM), and the coexistence of both signals an itinerant orbital ferrimagnetic state (OFiM). We then use a low-energy theory to interpret these regimes in terms of the valley structure of the two models.

Finally, we discuss how the orbital magnetic responses transform under spatial inversion ($P$), time reversal ($T$), and their combined operation ($PT$). This symmetry analysis is used as a consistency check for the orbital phases identified from $M_z$ and $M_z^s$. In particular, it shows how the net and staggered orbital responses parallel the transformation properties of collinear spin ferro-, antiferro-, and ferrimagnetic states, while retaining their itinerant orbital origin.

\begin{figure}[!h]
    \centering
    \includegraphics[width=1.0\linewidth]{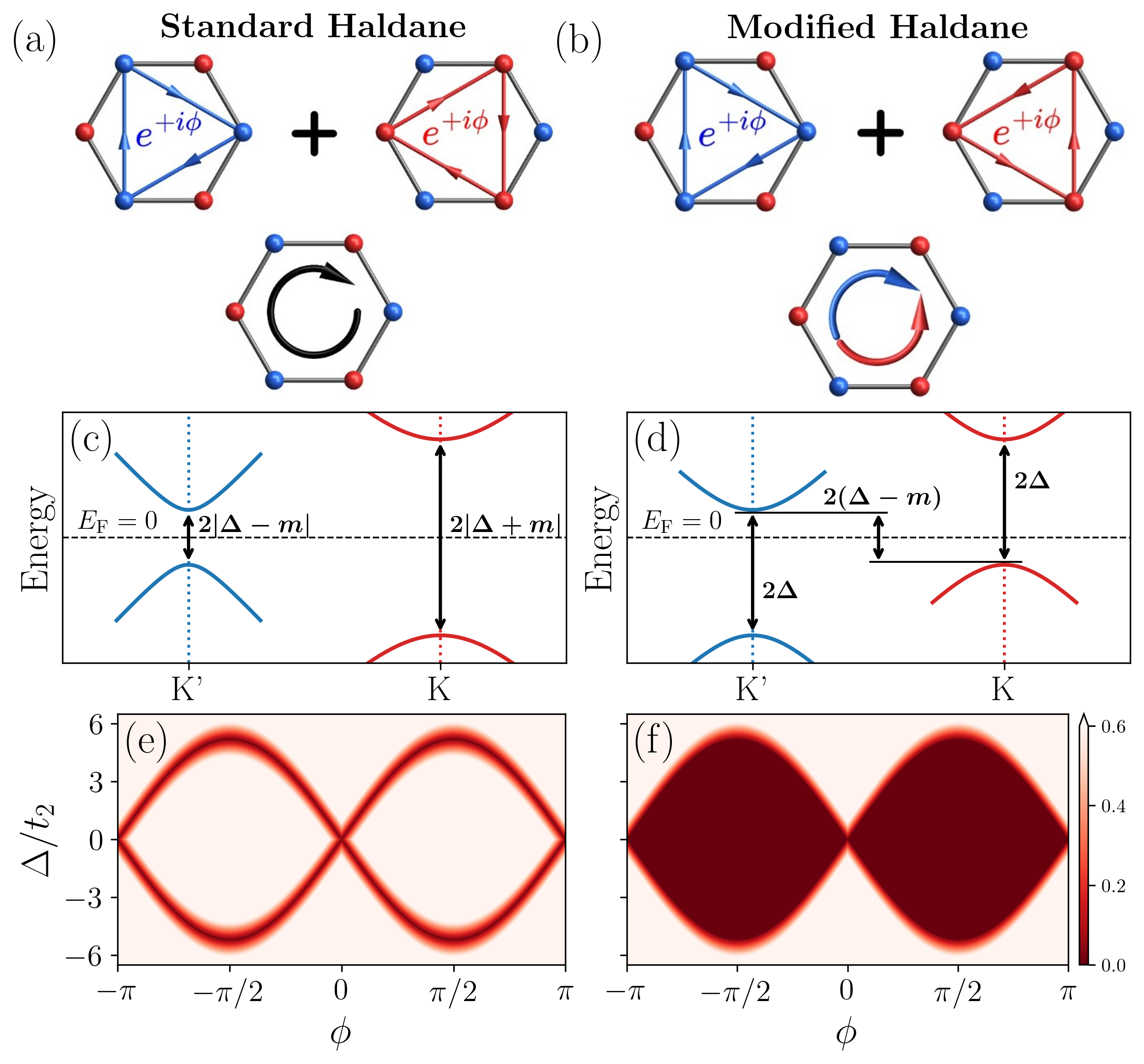}
   \caption{Convention for the positive phase $\phi$ in the $t_2$ term of the Hamiltonian in Eq. (\ref{Htot}) for the (a) standard and (b) modified Haldane models. We also show the schematic energy spectra near the valleys and the calculated bulk energy band gap (shown as a color gradient) across the parameter space $\Delta/t_2-\phi$ for the (c,e) standard and (d,f) modified Haldane models.
   } \label{fig1}
\end{figure}


{\bf Models and Methods:}
We consider the following tight-binding Hamiltonian for spinless fermions on a honeycomb lattice,
\begin{equation}
\mathcal{H} = t_1\sum_{\langle i, j \rangle}c^{\dagger}_{i}c_{ j} + \Delta\sum_i\xi_ic^{\dagger}_{i}c_{i} + t_2\sum_{\langle \langle i, j\rangle \rangle} e^{i \nu^{\alpha}_{ij} \phi} c^{\dagger}_{i} c_{j} + \text{H.c.}, \label{Htot}
\end{equation}
where $t_1$ is the hopping amplitude between nearest-neighbor sites $\langle i,j\rangle$. The term $\Delta$ represents the sublattice potential, with $\xi_i=+1$ ($-1$) for sites on sublattice $A$ ($B$). In the last term, $t_2$ is hopping between next-nearest-neighbor (NNN), accompanied by a complex phase. we consider two models that differ only in the choice of this phase. The first case corresponds to the standard Haldane model ($\alpha=$ H), in  which $\nu^{\rm H}_{ij}=+1$ ($-1$) for clockwise (counterclockwise) hoppings on sublattice $A$ ($B$) \cite{Haldane1988}, as ilustrated in Fig. \ref{fig1}(a). In the second case, we take  $\nu^{\rm mH}_{ij} = +1$ ($+1$) for clockwise (counterclockwise) hoppings in sublattice $A$ ($B$), as illustrated in Fig. \ref{fig1}(b), corresponding to a modified Haldane model ($\alpha=$ mH). Unless explicitly stated otherwise, we use $t_1=1 \text{eV}$ and $t_2=t_1/3$ for the numerical calculations. 

The modified Haldane model has received less attention than the standard one, despite exhibiting antichiral edge states \cite{Colomes2018}, perfect circular dichroism, and valley polarization \cite{Vila2019}. This model has been implemented in engineered photonic lattices \cite{Zhou2020}, and its spinful version appears in graphene/2H-transition metal dichalcogenide (TMD) heterostructures \cite{Frank2018, Srivastava2015} and in magnetic TMDs \cite{Tong2016}. 

The different complex-phase conventions strongly affect the electronic spectra and topological properties \cite{Vila2019,Lee2025,Mannai2020}. This becomes evident when Eq.~(\ref{Htot}) is transformed to ${\bf k}$-space [see Supplementary Material (SM)] and expanded around the valley points, yielding
\begin{eqnarray}
\mathcal{H}({\bf q})=\hbar v_{\rm F}(\tau_zq_x\sigma_x+q_y\sigma_y)+\Delta\sigma_z+m\sigma_{\alpha}\tau_z,
\label{Hlow}
\end{eqnarray}
where the Pauli matrices $\sigma_{0,x,y,z}$ act in the sublattice space $\{\ket{A},\ket{B}\}$, $\tau_z$ acts in the valley space $\{\ket{{\bf K}},\ket{{\bf K'}}\}$, and ${\bf q}={\bf k}-\tau{\bf K}$, with $\tau=\pm 1$, is the crystal momentum relative to the valley points ${\bf K}$ and ${\bf K'}=-{\bf K}$, respectively. $v_{\rm F}=3at_1/(2\hbar)$ is the Fermi velocity, and $m=3\sqrt{3}t_2\sin(\phi)$ is the mass generated by the NNN complex hoppings. In the last term, $\sigma_{\alpha}=\sigma_z$ for the standard Haldane model and $\sigma_{\alpha}=\sigma_0$ for the modified Haldane model.

In the standard Haldane model [see Fig.~\ref{fig1}(c),(e)], a \emph{direct band gap} $E_{\rm bg}=2|m-\Delta|$ opens at the valley points. For $\Delta=m$, the gap closes at one valley, producing the topological transition between the Chern insulating phase $(\Delta<m)$ and the trivial insulating phase $(\Delta>m)$. In the modified Haldane model [see Fig.~\ref{fig1}(d),(f)], the system remains metallic for $\Delta<m$, although \emph{local gaps} of size $E_{\rm lg}=2\Delta$ open at the energy-shifted Dirac cones. For $\Delta>m$, the system becomes a trivial insulator with an \emph{indirect band gap} $E_{\rm bg}=2(\Delta-m)$ \cite{Vila2019,Lee2025}.

We use these two models to investigate itinerant orbital ferromagnetic (OFM), antiferromagnetic (OAFM), and ferrimagnetic (OFiM) regimes. The calculations are performed in real space for finite samples with open boundary conditions. Our main numerical tool is the spectral approach introduced in Ref.~\cite{Vidarte2026}, based on a Chebyshev expansion of the spectral function associated with the orbital magnetization operator. Unlike the local marker, this method is not a strictly local construction: it gives the orbital magnetization of the full finite sample and therefore incorporates bulk and boundary contributions on the same footing. This is particularly useful for Haldane-type systems, where edge states and bulk orbital magnetization are closely connected.

As an independent check, we compare the spectral results with those obtained from the real-space local-marker theory of orbital magnetization \cite{Bianco2013}. Both methods provide the sublattice-resolved orbital magnetization contributions $M_A$ and $M_B$ for sublattices $A$ and $B$, from which we obtain $M_z$ and $M_z^s$. These two quantities are used below to identify OFM, OAFM, and OFiM responses.

\begin{figure}[h]
    \centering    \includegraphics[width=1.0\linewidth]{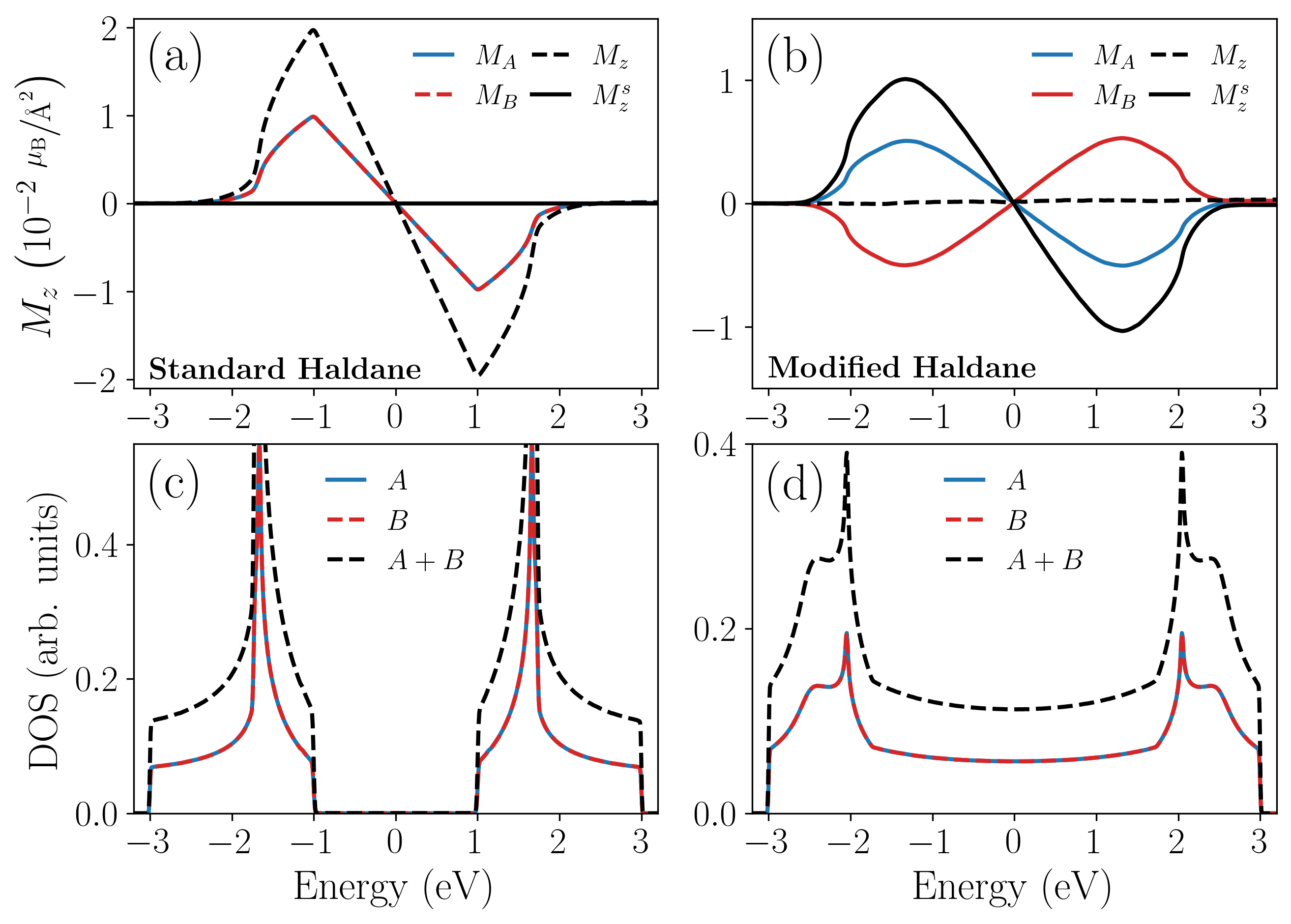}
   \caption{Sublattice-resolved ($M_A$ and $M_B$), total ($M_z$), and staggered ($M_z^s$) orbital magnetizations for the standard and modified Haldane models in panels (a) and (b), respectively, and the corresponding density of states in panels (c) and (d), for $\phi=\pi/2$ and $\Delta=0$.} 
   \label{fig2}
\end{figure}

{\bf Results:}
We begin with the sublattice-balanced case, $\Delta=0$, taking $\phi=\pi/2$ in both models. Figure~\ref{fig2} shows the corresponding sublattice-resolved density of states (DOS) and orbital magnetization. In the standard Haldane model, the two sublattices contribute equally to the orbital magnetization, $M_A=M_B$, over the full range of Fermi energies. Consequently, the net orbital magnetization is finite while the staggered component vanishes, $M_z\neq0$ and $M_z^s=0$, identifying this regime as an itinerant orbital ferromagnet (OFM).

The modified Haldane model displays the complementary behavior. In this case, the sublattice-resolved orbital magnetizations satisfy $M_A=-M_B$, and then the net magnetization cancels while the staggered component remains finite, $M_z=0$ and $M_z^s\neq0$. This corresponds to an itinerant orbital antiferromagnet (OAFM). As discussed below, this distinction is consistent with the $PT$ symmetry of the modified Haldane model at $\Delta=0$. 

\begin{figure}[h]
    \centering
    \includegraphics[width=1.0\linewidth]{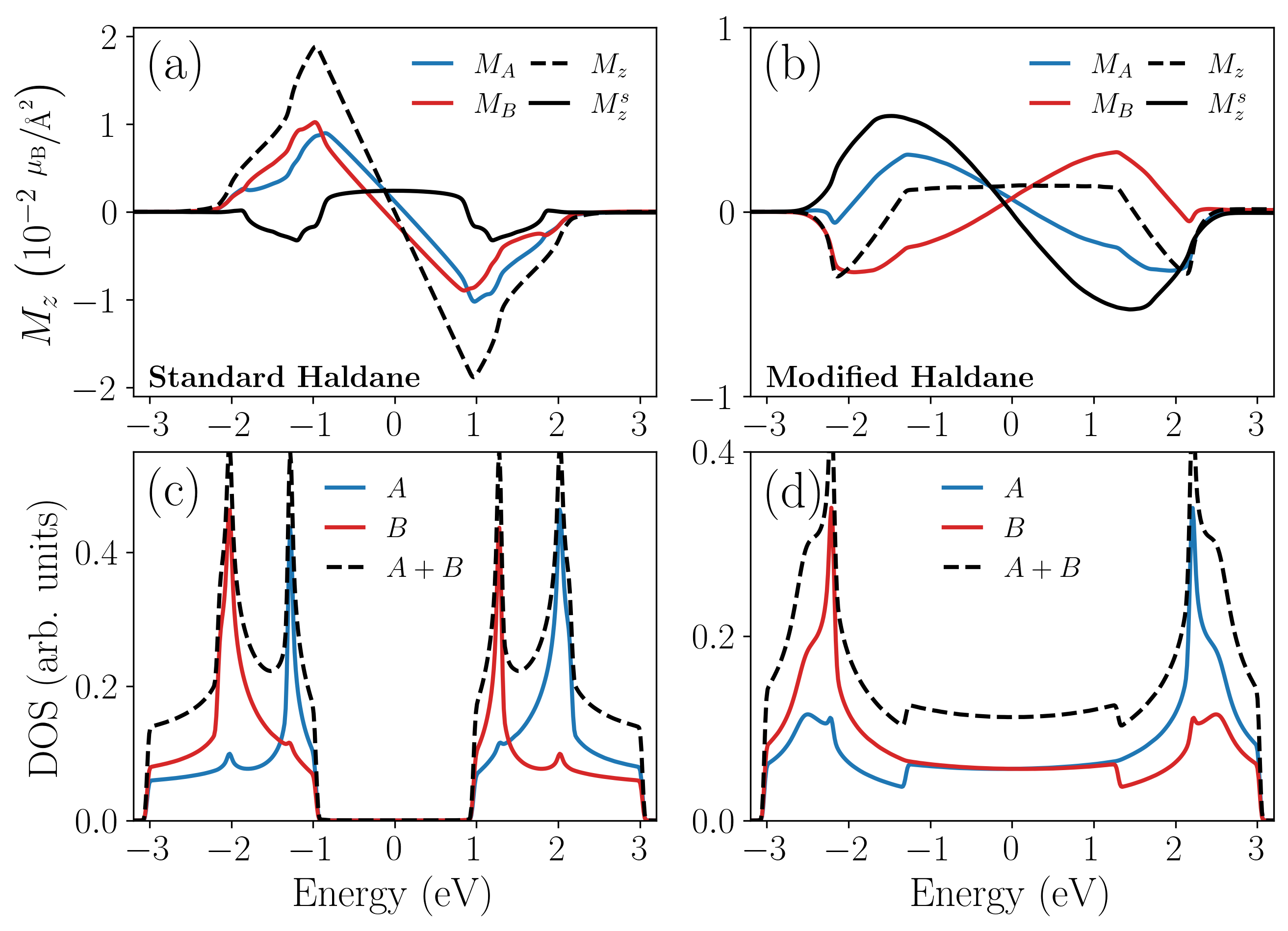}
   \caption{Sublattice-resolved ($M_A$ and $M_B$), total ($M_z$), and staggered ($M_z^s$) orbital magnetizations for the standard and modified Haldane models in panels (a) and (b), respectively, and the corresponding density of states in panels (c) and (d), for $\phi=\pi/2$, and $\Delta=0.75\sqrt{3}t_2$.} 
   \label{fig3}
\end{figure}

\begin{figure}[h]
    \centering
    \includegraphics[width=1.0\linewidth]{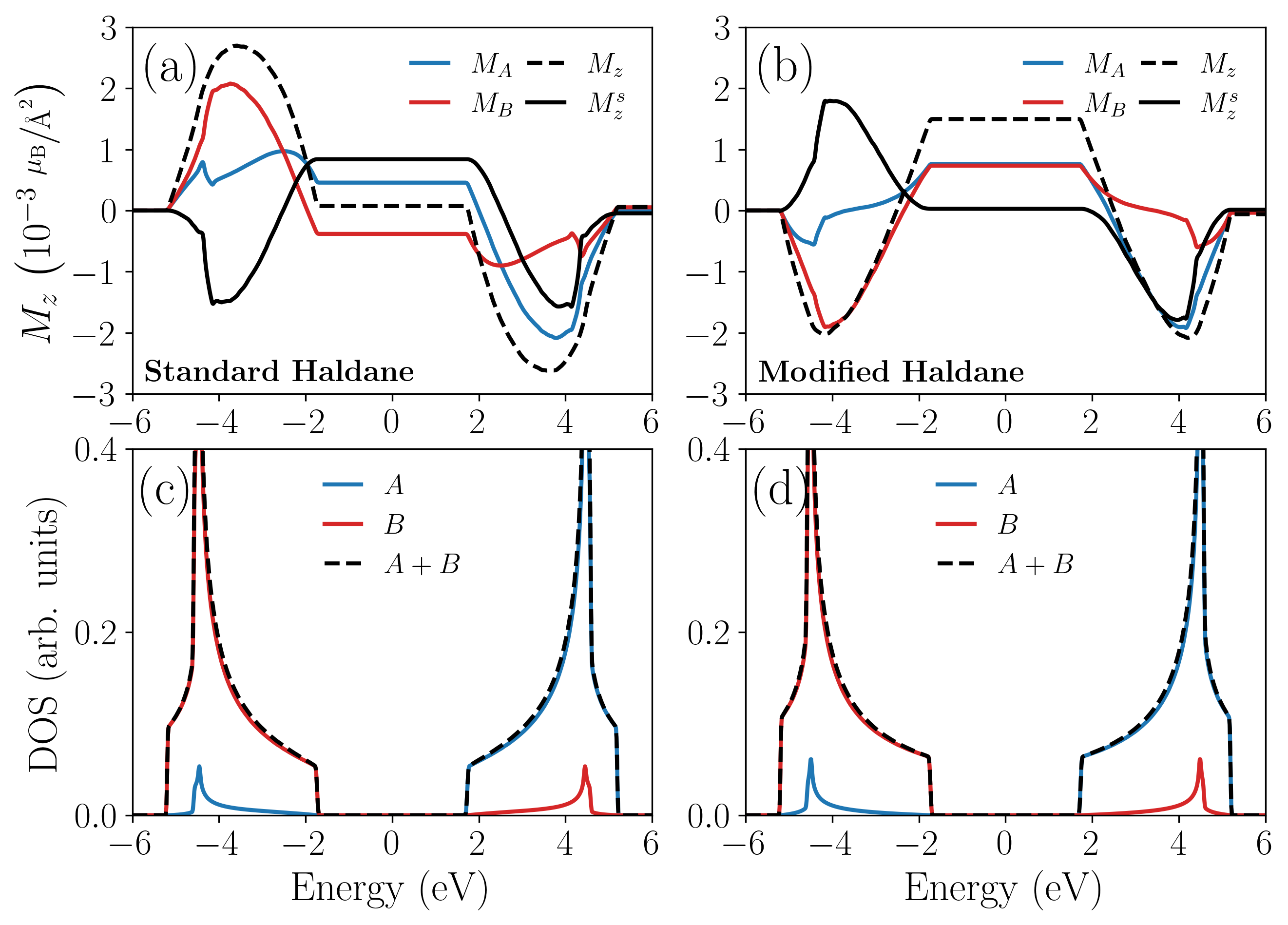}
   \caption{Sublattice-resolved ($M_A$ and $M_B$), total ($M_z$), and staggered ($M_z^s$) orbital magnetizations for the standard and modified Haldane models in panels (a) and (b), respectively, and the corresponding density of states in panels (c) and (d), for $\phi=\pi/2$, and $\Delta=6\sqrt{3}t_2$.
   } \label{fig4}
\end{figure}

We next introduce a finite sublattice potential in the regime $\Delta<m$, as shown in Fig.~\ref{fig3}. The sublattice imbalance makes the DOS on the two sublattices inequivalent and removes the exact relations found at $\Delta=0$. In the standard Haldane model, $M_A\neq M_B$, while in the modified Haldane model, $M_A\neq -M_B$. Both the net and staggered components of the orbital magnetization are therefore finite, $M_z\neq0$ and $M_z^s\neq0$, identifying both systems as itinerant orbital ferrimagnets (OFiMs) throughout the energy range shown.

The most revealing regime occurs for $\Delta>m$ at fixed $\phi=\pi/2$, as shown in Fig.~\ref{fig4}. In the standard Haldane model, this condition places the system in the topologically trivial insulating phase. In the modified Haldane model, it opens an indirect band gap. When the Fermi energy crosses states with finite DOS, both models again show finite net and staggered orbital magnetizations, as in the OFiM regime discussed above. However, when the Fermi energy lies inside the insulating gap, the orbital magnetization develops a plateau whose character depends on the model. Surprisingly, in the standard Haldane model, the net orbital magnetization cancels while the staggered component remains finite, $M_z=0$ and $M_z^s\neq0$, corresponding to an OAFM [Fig.~\ref{fig4}(a)]. In the modified Haldane model, the opposite occurs: $M_z\neq0$ and $M_z^s=0$, corresponding to an OFM [Fig.~\ref{fig4}(b)].

Thus, the trivial insulating gap reveals a reversal of the orbital character of the two models relative to the sublattice-balanced case. The standard Haldane model, which is orbital ferromagnetic at $\Delta=0$, becomes orbital antiferromagnetic in the insulating plateau. Conversely, the modified Haldane model, which is orbital antiferromagnetic at $\Delta=0$, develops an orbital ferromagnetic plateau. This interchange is the central feature that the low-energy analysis below explains in terms of the valley structure of the two Hamiltonians.

{\bf Low energy analysis:} 
The insulating plateaus in Fig.~\ref{fig4} reveal the most intriguing feature of the real-space results.  In the trivial insulating regime, $\Delta>m$ and $\phi=\pi/2$, the standard Haldane model displays a purely staggered orbital magnetization, whereas the modified Haldane model displays a purely net orbital magnetization.  We now use the low-energy Hamiltonian of Eq.~(\ref{Hlow}) to explain the origin of this interchange.  For the modified Haldane model, the net plateau can be obtained directly from the modern theory of orbital magnetization in momentum space.  For the standard Haldane model, where the relevant response is staggered, we introduce sublattice projectors in the same modern-theory expression to extract the contributions associated with the two sublattices.

We begin by describing the OFM in the insulating region of the modified Haldane model shown in Fig. \ref{fig4} (b). The {\bf k}-space expression for orbital magnetization can be written as \cite{Xiao2005}: $M_z=\sum_{n}\int_{\epsilon_{n{\bf k}}\le E_F}\frac{d{\bf k}}{(2\pi)^2} \left\{m_{n{\bf k}}+(e/c\hbar)(E_{F}-\epsilon_{n{\bf k}})\Omega_{n{\bf k}}\right\}$. Computing the magnetic moment of Bloch states $m_{n{\bf k}}=-i(e/2c\hbar)\bra{\boldsymbol{\nabla}_{\bf k}u_{n{\bf k}}}\boldsymbol{\times}(\hat{H}_{n{\bf k}}-\epsilon_{n{\bf k}}\mathbb{1})\ket{\boldsymbol{\nabla}_{\bf k}u_{n{\bf k}}}$ and Berry curvature $\Omega_{n{\bf k}}=i \bra{\boldsymbol{\nabla}_{\bf k}u_{n{\bf k}}}\boldsymbol{\times}\ket{\boldsymbol{\nabla}_{\bf k}u_{n{\bf k}}}$ using low-energy Hamiltonian of Eq. (\ref{Hlow}) with $\sigma_{\alpha} \rightarrow \sigma_0$, one obtains for the valence band ($n=v$): $\Omega_{v{\bf q}}=\tau\hbar^2v_{\rm F}^2\Delta/(2\varepsilon_q^3)$ and $m_{v{\bf q}}=-\tau (e/c\hbar)\hbar^2v^2_{\rm F}\Delta/(2\varepsilon^2_{q})$, where $\varepsilon_q=\sqrt{\hbar^2v^2_{\rm F}q^2+\Delta^2}$. When $\Delta > m$ and $E_{\rm F}$ lies within the insulating band gap, the contributions to $M_z$ from the orbital magnetic moment cancel when summed over the valleys: $\sum_{\tau}\int\frac{d{\bf q}}{(2\pi)^2}m_{v{\bf q}}=0$. Nevertheless, since the valence-band eigenenergy can be written as $\epsilon_{v{\bf q}}=\tau m-\varepsilon_q$, the Berry-curvature term in the ${\bf k}$-expression for the $M_z$ reads: $(e/c\hbar)(E_{\rm F}-\epsilon_{v{\bf q}})\Omega_{v{\bf q}}=-(e/c\hbar)m\tau\Omega_{v{\bf q}}+(e/c\hbar)(E_{\rm F}-\varepsilon_q)\Omega_{v{\bf q}}$. While the second term again cancels out between the valleys [$\sum_{\tau}\int\frac{d{\bf q}}{(2\pi)^2}\Omega_{v{\bf q}}=0$], the first term gives rise to an orbital magnetization plateau reported in Fig. \ref{fig4} (b): $M_z=(e/c\hbar)m(2\pi)^{-1}\int^0_{\infty}\hbar^2 v^2_{\rm F}\Delta\varepsilon_q^{-3}qdq=(e/c\hbar)m(2\pi)^{-1}$.  
 
Using the low-energy theory of Eq. (\ref{Hlow}) with $\sigma_{\alpha}\rightarrow\sigma_{z}$, we now explore the origin of the OAFM plateau that appears in the standard Haldane model for $\phi=\pi/2$ and $\Delta>m$. To this end, we define the sublattice projector $\mathcal{S}_{\sigma}=\ket{\sigma}\bra{\sigma}$, where $\sigma=A,B$, as well as the band projectors $\mathcal{P}_{n}=\ket{u_{n{\bf q}}}\bra{u_{n{\bf q}}}$, where $n=c,v$ for the conduction and valence bands. The sublattice-resolved weight for each band is given by $w_{n\sigma{\bf q}}\equiv \bra{u_{n{\bf q}}}\mathcal{S}_{\sigma}\ket{u_{n{\bf q}}}=\text{Tr}\left[\mathcal{S}_{\sigma} \mathcal{P}_{n}\right]$. Using these definitions, we project the {\bf k}-space expression for orbital magnetization onto the $A$ and $B$ sublattices. This sublattice projection is used here as a low-energy description of the non-inverted trivial insulating regime, allowing us to interpret the staggered plateau obtained in the real-space calculations. The sublattice-projected orbital magnetization can then be written as
$M_{\sigma}=\frac{e}{\hbar c}\sum_{\tau=\pm 1} \int\frac{d{\bf q}}{(2\pi)^2}\mathcal{I}_{\sigma{\bf q}}$, 
where 
$\mathcal{I}_{\sigma{\bf q}}=\left[\tilde{\epsilon}_{c{\bf q}}w_{v\sigma{\bf q}}+(\tilde{\epsilon}_{v{\bf q}}-2E_{\rm F})w_{c\sigma{\bf q}} \right]\Lambda_{\bf q}$. 
Here, $\tilde{\epsilon}_{c(v){\bf q}}=\pm \tilde{\varepsilon}_{q}$, with 
$\tilde{\varepsilon}_{q}=\sqrt{\hbar^2v^2_{\rm F}q^2+(\Delta-\tau m)^2}$, and 
$\Lambda_{{\bf q}}\equiv \text{Im}\left[ \braket{\partial_{x}u_{v{\bf q}}}{u_{c{\bf q}}} \braket{u_{c{\bf q}}}{\partial_{y}u_{v{\bf q}}}\right]$. 
In a two-band model, this quantity is related to the valence-band Berry curvature by 
$\Lambda_{\bf q}=-\tilde{\Omega}_{v{\bf q}}/2$.

The essential point is that the sublattice projection separates the contributions that cancel in the net magnetization from those that survive in the staggered channel. Computing the weights $w_{n\sigma{\bf q}}$ explicitly gives
$\mathcal{I}_{A(B){\bf q}}=\pm\frac{1}{2}\left[\Delta-\tau m \pm E_{\rm F}\left(1\pm (\Delta-\tau m)/\tilde{\varepsilon}_q \right)\right]\tilde{\Omega}_{v{\bf q}}$.
When forming the staggered combination, $\mathcal{I}_{A{\bf q}}-\mathcal{I}_{B{\bf q}}$, the terms linear in $E_{\rm F}$ cancel after summing over the valleys. The remaining $E_{\rm F}$-independent contribution gives
$M_z^s=\frac{e}{\hbar c}\sum_{\tau}\int^0_{\infty}\frac{qdq}{(2\pi)}\left[\mathcal{I}_{A{\bf q}}-\mathcal{I}_{B{\bf q}}\right]=(e/c\hbar)m(2\pi)^{-1}$.
Thus, the staggered orbital magnetization remains finite throughout the insulating plateau.

By contrast, the net orbital magnetization is obtained from the sum of the two sublattice contributions. In this case, one finds
$\mathcal{I}_{A{\bf q}}+\mathcal{I}_{B{\bf q}}=E_{\rm F}\tilde{\Omega}_{v{\bf q}}$.
In the trivial insulating phase, the total Berry curvature integrated over the two valleys vanishes,
$\sum_{\tau}\int\frac{d{\bf q}}{(2\pi)^2}\tilde{\Omega}_{v{\bf q}}=0$, and therefore $M_z=0$. 
This explains why the standard Haldane model develops an OAFM plateau in the trivial insulating regime of Fig.~\ref{fig4}(a): the net orbital magnetization cancels between valleys, while the sublattice-staggered contribution does not. Further details of the calculation are provided in the SM.

The insulating OAFM (OFM) plateau of the standard (modified) Haldane model in Fig.~\ref{fig4} relies on particle-hole symmetry. For $\phi\neq\pi/2$, the trivial insulating phase generally develops both $M_z$ and $M_z^s$, so that the two models become OFiMs (see SM).

\begin{table}[h]
	\centering
	\begin{tabular}{||c c |c c c c c | c c  c c| c c ||} 
		\hline
		Models & & $P$ & &  $T$ & &  $P.T$ && $M_z$ && $M^{s}_z$&& order\\ [0.5ex] 
		\hline
		\hline
		(I) Standard ($\Delta=0$) & & $+1$ & & $-1$ & & $-1$ & & Y & & N && OFM \\
		(II) Modified ($\Delta=0$) & & $-1$ & & $-1$ & & $+1$ & & N & & Y && OAFM \\
            (III) Standard ($\Delta\neq0$) & & $\times$ & & $\times$ & & $-1$ & & Y & & Y && OFiM\\
            (IV) Modified ($\Delta\neq0$) & & $-1$ & & $\times$ & & $\times$ & & Y & & Y && OFiM\\
        [0.5ex] 
		\hline
	\end{tabular}
	\caption{Classification of the standard and modified Haldane models with respect to $P$, $T$, and $PT$ symmetries at valleys ${\bf K}$ and ${\bf K'}$. $+1$ and $-1$ indicate that the model is even (preserves) and odd (breaks) under the symmetry at the valleys, respectively. The symbol $\times$ indicates that the model breaks a given symmetry without a well-defined parity. In the fourth and fifth columns, Yes (Y) or No (N) indicates whether the model exhibits finite $M_z$ and $M^s_z$. The last column indicates the orbital order.} 
	\label{table1}
\end{table}

{\bf $P$, $T$, and $PT$ symmetries:}
We now discuss how the orbital magnetic regimes identified above transform under spatial inversion ($P$), time reversal ($T$), and their product ($PT$). This provides a symmetry consistency check with the usual spin analogues. For a collinear spin order on a bipartite lattice, with inversion exchanging the two sublattices, a ferromagnet preserves $P$ and breaks $T$ and $PT$, an antiferromagnet breaks $P$ and $T$ but preserves $PT$, while a ferrimagnet breaks all three symmetries.

We verify the same pattern in the low-energy Hamiltonian of Eq.~(\ref{Hlow}) by considering the valley points ${\bf K}$ and ${\bf K'}$, where the Dirac term vanishes (${\bf q}\rightarrow0$). In the standard Haldane model, the relevant term is $m\sigma_z\tau_z$, which transforms as $P:\sigma_z\tau_z\rightarrow\sigma_z\tau_z$, $T:\sigma_z\tau_z\rightarrow-\sigma_z\tau_z$, and $PT:\sigma_z\tau_z\rightarrow-\sigma_z\tau_z$ (see SM). Thus, for $\Delta=0$, the standard Haldane model has the symmetry character of an OFM, consistent with $M_z\neq0$ and $M_z^s=0$ [Table~\ref{table1}, case (I)]. In the modified Haldane model, the corresponding term is $m\sigma_0\tau_z$, which transforms as $P:\sigma_0\tau_z\rightarrow-\sigma_0\tau_z$, $T:\sigma_0\tau_z\rightarrow-\sigma_0\tau_z$, and $PT:\sigma_0\tau_z\rightarrow\sigma_0\tau_z$. Therefore, at $\Delta=0$, the modified Haldane model has the symmetry character of an OAFM, in agreement with $M_z=0$ and $M_z^s\neq0$ [Table~\ref{table1}, case (II)].

Finally, the sublattice potential transforms as $P:\sigma_z\rightarrow-\sigma_z$, $T:\sigma_z\rightarrow\sigma_z$, and $PT:\sigma_z\rightarrow-\sigma_z$. When $\Delta\neq0$, this term removes the symmetry protection of the pure OFM and OAFM limits, and the generic situation is the coexistence of net and staggered orbital magnetizations, analogous to a ferrimagnet [Table~\ref{table1}, cases (III) and (IV)].

{\bf Final Remarks and Conclusions:}
We have shown that itinerant orbital magnetization can display ferro-, antiferro-, and ferrimagnetic character when resolved into sublattice contributions. Using the standard and modified Haldane models as minimal examples, we identified regimes with net, staggered, and mixed orbital magnetic responses. These states do not rely on assigning localized orbital moments to individual sites, but arise from the itinerant circulation of electronic states and are captured by real-space formulations of orbital magnetization.

The low-energy analysis shows that the different orbital responses originate from distinct valley mechanisms: valley-dependent Dirac masses in the standard Haldane model and valley-dependent energy shifts in the modified Haldane model. Their transformation properties under $P$, $T$, and $PT$ further parallel those of collinear spin ferro-, antiferro-, and ferrimagnets.

The analogy with spin magnetic order suggests possible directions for orbitronics. Antiferromagnetic and ferrimagnetic materials are actively explored in spintronics because of their fast dynamics, robustness, and, in ferrimagnets, controllable net magnetic response \cite{Kim2021,Baltz2018,Jungwirth2016}. In the orbital context, distinguishing net and staggered orbital magnetization may provide a useful route to characterize orbital angular momentum in itinerant systems and to explore orbital order beyond the atom-centered picture.

\begin{acknowledgments}
TGR acknowledges the  FCT - Fundação para a Ciência e Tecnologia for financial support (Grant No. 2023.11755.PEX and DOI identifier https://doi.org/10.54499/2023.11755.PEX), FAPERJ "Cientistas do Nosso Estado" and CNPq (Grant No 305013/2024-6). TGR and KJUV acknowledge support from the EIC Pathfinder OPEN grant 101129641 “OBELIX”. TPC acknowledges CNPq (Grant No. 305647/2024-5). A.L. thanks
FAPERJ (Grant No. E-26/200.569/2023 and SEI-260003/021951/2025).
\end{acknowledgments}

\begin{thebibliography}{50}%
\makeatletter
\providecommand \@ifxundefined [1]{%
 \@ifx{#1\undefined}
}%
\providecommand \@ifnum [1]{%
 \ifnum #1\expandafter \@firstoftwo
 \else \expandafter \@secondoftwo
 \fi
}%
\providecommand \@ifx [1]{%
 \ifx #1\expandafter \@firstoftwo
 \else \expandafter \@secondoftwo
 \fi
}%
\providecommand \natexlab [1]{#1}%
\providecommand \enquote  [1]{``#1''}%
\providecommand \bibnamefont  [1]{#1}%
\providecommand \bibfnamefont [1]{#1}%
\providecommand \citenamefont [1]{#1}%
\providecommand \href@noop [0]{\@secondoftwo}%
\providecommand \href [0]{\begingroup \@sanitize@url \@href}%
\providecommand \@href[1]{\@@startlink{#1}\@@href}%
\providecommand \@@href[1]{\endgroup#1\@@endlink}%
\providecommand \@sanitize@url [0]{\catcode `\\12\catcode `\$12\catcode `\&12\catcode `\#12\catcode `\^12\catcode `\_12\catcode `\%12\relax}%
\providecommand \@@startlink[1]{}%
\providecommand \@@endlink[0]{}%
\providecommand \url  [0]{\begingroup\@sanitize@url \@url }%
\providecommand \@url [1]{\endgroup\@href {#1}{\urlprefix }}%
\providecommand \urlprefix  [0]{URL }%
\providecommand \Eprint [0]{\href }%
\providecommand \doibase [0]{https://doi.org/}%
\providecommand \selectlanguage [0]{\@gobble}%
\providecommand \bibinfo  [0]{\@secondoftwo}%
\providecommand \bibfield  [0]{\@secondoftwo}%
\providecommand \translation [1]{[#1]}%
\providecommand \BibitemOpen [0]{}%
\providecommand \bibitemStop [0]{}%
\providecommand \bibitemNoStop [0]{.\EOS\space}%
\providecommand \EOS [0]{\spacefactor3000\relax}%
\providecommand \BibitemShut  [1]{\csname bibitem#1\endcsname}%
\let\auto@bib@innerbib\@empty
\bibitem [{\citenamefont {Go}\ \emph {et~al.}(2021)\citenamefont {Go}, \citenamefont {Jo}, \citenamefont {Gao}, \citenamefont {Ando}, \citenamefont {Bl\"ugel}, \citenamefont {Lee},\ and\ \citenamefont {Mokrousov}}]{Go2021}%
  \BibitemOpen
  \bibfield  {author} {\bibinfo {author} {\bibfnamefont {D.}~\bibnamefont {Go}}, \bibinfo {author} {\bibfnamefont {D.}~\bibnamefont {Jo}}, \bibinfo {author} {\bibfnamefont {T.}~\bibnamefont {Gao}}, \bibinfo {author} {\bibfnamefont {K.}~\bibnamefont {Ando}}, \bibinfo {author} {\bibfnamefont {S.}~\bibnamefont {Bl\"ugel}}, \bibinfo {author} {\bibfnamefont {H.-W.}\ \bibnamefont {Lee}},\ and\ \bibinfo {author} {\bibfnamefont {Y.}~\bibnamefont {Mokrousov}},\ }\bibfield  {title} {\bibinfo {title} {Orbital rashba effect in a surface-oxidized cu film},\ }\href {https://doi.org/10.1103/PhysRevB.103.L121113} {\bibfield  {journal} {\bibinfo  {journal} {Phys. Rev. B}\ }\textbf {\bibinfo {volume} {103}},\ \bibinfo {pages} {L121113} (\bibinfo {year} {2021})}\BibitemShut {NoStop}%
\bibitem [{\citenamefont {Jo}\ \emph {et~al.}(2024)\citenamefont {Jo}, \citenamefont {Go}, \citenamefont {Choi},\ and\ \citenamefont {Lee}}]{Jo2024}%
  \BibitemOpen
  \bibfield  {author} {\bibinfo {author} {\bibfnamefont {D.}~\bibnamefont {Jo}}, \bibinfo {author} {\bibfnamefont {D.}~\bibnamefont {Go}}, \bibinfo {author} {\bibfnamefont {G.-M.}\ \bibnamefont {Choi}},\ and\ \bibinfo {author} {\bibfnamefont {H.-W.}\ \bibnamefont {Lee}},\ }\bibfield  {title} {\bibinfo {title} {Spintronics meets orbitronics: Emergence of orbital angular momentum in solids},\ }\bibfield  {journal} {\bibinfo  {journal} {npj Spintronics}\ }\textbf {\bibinfo {volume} {2}},\ \href {https://doi.org/10.1038/s44306-024-00023-6} {10.1038/s44306-024-00023-6} (\bibinfo {year} {2024})\BibitemShut {NoStop}%
\bibitem [{\citenamefont {Atencia}\ \emph {et~al.}(2024)\citenamefont {Atencia}, \citenamefont {Agarwal},\ and\ \citenamefont {Culcer}}]{Atencia2024}%
  \BibitemOpen
  \bibfield  {author} {\bibinfo {author} {\bibfnamefont {R.~B.}\ \bibnamefont {Atencia}}, \bibinfo {author} {\bibfnamefont {A.}~\bibnamefont {Agarwal}},\ and\ \bibinfo {author} {\bibfnamefont {D.}~\bibnamefont {Culcer}},\ }\bibfield  {title} {\bibinfo {title} {Orbital angular momentum of bloch electrons: equilibrium formulation, magneto-electric phenomena, and the orbital hall effect},\ }\href {https://doi.org/10.1080/23746149.2024.2371972} {\bibfield  {journal} {\bibinfo  {journal} {Advances in Physics: X}\ }\textbf {\bibinfo {volume} {9}},\ \bibinfo {pages} {2371972} (\bibinfo {year} {2024})},\ \Eprint {https://arxiv.org/abs/https://doi.org/10.1080/23746149.2024.2371972} {https://doi.org/10.1080/23746149.2024.2371972} \BibitemShut {NoStop}%
\bibitem [{\citenamefont {Cysne}\ \emph {et~al.}(2025)\citenamefont {Cysne}, \citenamefont {Canonico}, \citenamefont {Costa}, \citenamefont {Muniz},\ and\ \citenamefont {Rappoport}}]{Cysne2025}%
  \BibitemOpen
  \bibfield  {author} {\bibinfo {author} {\bibfnamefont {T.~P.}\ \bibnamefont {Cysne}}, \bibinfo {author} {\bibfnamefont {L.~M.}\ \bibnamefont {Canonico}}, \bibinfo {author} {\bibfnamefont {M.}~\bibnamefont {Costa}}, \bibinfo {author} {\bibfnamefont {R.~B.}\ \bibnamefont {Muniz}},\ and\ \bibinfo {author} {\bibfnamefont {T.~G.}\ \bibnamefont {Rappoport}},\ }\bibfield  {title} {\bibinfo {title} {Orbitronics in two-dimensional materials},\ }\href {https://doi.org/10.1038/s44306-025-00103-1} {\bibfield  {journal} {\bibinfo  {journal} {npj Spintronics}\ }\textbf {\bibinfo {volume} {3}},\ \bibinfo {pages} {39} (\bibinfo {year} {2025})}\BibitemShut {NoStop}%
\bibitem [{\citenamefont {Ando}(2025)}]{Ando2025}%
  \BibitemOpen
  \bibfield  {author} {\bibinfo {author} {\bibfnamefont {K.}~\bibnamefont {Ando}},\ }\bibfield  {title} {\bibinfo {title} {Orbitronics: Harnessing orbital currents in solid-state devices},\ }\bibfield  {journal} {\bibinfo  {journal} {Journal of the Physical Society of Japan}\ }\textbf {\bibinfo {volume} {94}},\ \href {https://doi.org/10.7566/jpsj.94.092001} {10.7566/jpsj.94.092001} (\bibinfo {year} {2025})\BibitemShut {NoStop}%
\bibitem [{\citenamefont {Fukami}\ \emph {et~al.}(2025)\citenamefont {Fukami}, \citenamefont {Lee},\ and\ \citenamefont {Kl\"{a}ui}}]{Fukami2025}%
  \BibitemOpen
  \bibfield  {author} {\bibinfo {author} {\bibfnamefont {S.}~\bibnamefont {Fukami}}, \bibinfo {author} {\bibfnamefont {K.-J.}\ \bibnamefont {Lee}},\ and\ \bibinfo {author} {\bibfnamefont {M.}~\bibnamefont {Kl\"{a}ui}},\ }\bibfield  {title} {\bibinfo {title} {Challenges and opportunities in orbitronics},\ }\bibfield  {journal} {\bibinfo  {journal} {Nature Physics}\ }\href {https://doi.org/10.1038/s41567-025-03143-w} {10.1038/s41567-025-03143-w} (\bibinfo {year} {2025})\BibitemShut {NoStop}%
\bibitem [{\citenamefont {Wang}\ \emph {et~al.}(2024)\citenamefont {Wang}, \citenamefont {Chen}, \citenamefont {Yang}, \citenamefont {Hu}, \citenamefont {Li}, \citenamefont {Wang}, \citenamefont {Zhang},\ and\ \citenamefont {Jiang}}]{Wang2024}%
  \BibitemOpen
  \bibfield  {author} {\bibinfo {author} {\bibfnamefont {P.}~\bibnamefont {Wang}}, \bibinfo {author} {\bibfnamefont {F.}~\bibnamefont {Chen}}, \bibinfo {author} {\bibfnamefont {Y.}~\bibnamefont {Yang}}, \bibinfo {author} {\bibfnamefont {S.}~\bibnamefont {Hu}}, \bibinfo {author} {\bibfnamefont {Y.}~\bibnamefont {Li}}, \bibinfo {author} {\bibfnamefont {W.}~\bibnamefont {Wang}}, \bibinfo {author} {\bibfnamefont {D.}~\bibnamefont {Zhang}},\ and\ \bibinfo {author} {\bibfnamefont {Y.}~\bibnamefont {Jiang}},\ }\bibfield  {title} {\bibinfo {title} {Orbitronics: Mechanisms, materials and devices},\ }\bibfield  {journal} {\bibinfo  {journal} {Advanced Electronic Materials}\ }\textbf {\bibinfo {volume} {11}},\ \href {https://doi.org/10.1002/aelm.202400554} {10.1002/aelm.202400554} (\bibinfo {year} {2024})\BibitemShut {NoStop}%
\bibitem [{\citenamefont {Choi}\ \emph {et~al.}(2023)\citenamefont {Choi}, \citenamefont {Jo}, \citenamefont {Ko}, \citenamefont {Go}, \citenamefont {Kim}, \citenamefont {Park}, \citenamefont {Kim}, \citenamefont {Min}, \citenamefont {Choi},\ and\ \citenamefont {Lee}}]{Choi2023}%
  \BibitemOpen
  \bibfield  {author} {\bibinfo {author} {\bibfnamefont {Y.-G.}\ \bibnamefont {Choi}}, \bibinfo {author} {\bibfnamefont {D.}~\bibnamefont {Jo}}, \bibinfo {author} {\bibfnamefont {K.-H.}\ \bibnamefont {Ko}}, \bibinfo {author} {\bibfnamefont {D.}~\bibnamefont {Go}}, \bibinfo {author} {\bibfnamefont {K.-H.}\ \bibnamefont {Kim}}, \bibinfo {author} {\bibfnamefont {H.~G.}\ \bibnamefont {Park}}, \bibinfo {author} {\bibfnamefont {C.}~\bibnamefont {Kim}}, \bibinfo {author} {\bibfnamefont {B.-C.}\ \bibnamefont {Min}}, \bibinfo {author} {\bibfnamefont {G.-M.}\ \bibnamefont {Choi}},\ and\ \bibinfo {author} {\bibfnamefont {H.-W.}\ \bibnamefont {Lee}},\ }\bibfield  {title} {\bibinfo {title} {Observation of the orbital hall effect in a light metal ti},\ }\href {https://doi.org/10.1038/s41586-023-06101-9} {\bibfield  {journal} {\bibinfo  {journal} {Nature}\ }\textbf {\bibinfo {volume} {619}},\ \bibinfo {pages} {52} (\bibinfo {year} {2023})}\BibitemShut {NoStop}%
\bibitem [{\citenamefont {Bernevig}\ \emph {et~al.}(2005)\citenamefont {Bernevig}, \citenamefont {Hughes},\ and\ \citenamefont {Zhang}}]{Bernevig2005}%
  \BibitemOpen
  \bibfield  {author} {\bibinfo {author} {\bibfnamefont {B.~A.}\ \bibnamefont {Bernevig}}, \bibinfo {author} {\bibfnamefont {T.~L.}\ \bibnamefont {Hughes}},\ and\ \bibinfo {author} {\bibfnamefont {S.-C.}\ \bibnamefont {Zhang}},\ }\bibfield  {title} {\bibinfo {title} {Orbitronics: The intrinsic orbital current in $p$-doped silicon},\ }\href {https://doi.org/10.1103/PhysRevLett.95.066601} {\bibfield  {journal} {\bibinfo  {journal} {Phys. Rev. Lett.}\ }\textbf {\bibinfo {volume} {95}},\ \bibinfo {pages} {066601} (\bibinfo {year} {2005})}\BibitemShut {NoStop}%
\bibitem [{\citenamefont {Cysne}\ \emph {et~al.}(2021)\citenamefont {Cysne}, \citenamefont {Costa}, \citenamefont {Canonico}, \citenamefont {Nardelli}, \citenamefont {Muniz},\ and\ \citenamefont {Rappoport}}]{Cysne2021}%
  \BibitemOpen
  \bibfield  {author} {\bibinfo {author} {\bibfnamefont {T.~P.}\ \bibnamefont {Cysne}}, \bibinfo {author} {\bibfnamefont {M.}~\bibnamefont {Costa}}, \bibinfo {author} {\bibfnamefont {L.~M.}\ \bibnamefont {Canonico}}, \bibinfo {author} {\bibfnamefont {M.~B.}\ \bibnamefont {Nardelli}}, \bibinfo {author} {\bibfnamefont {R.~B.}\ \bibnamefont {Muniz}},\ and\ \bibinfo {author} {\bibfnamefont {T.~G.}\ \bibnamefont {Rappoport}},\ }\bibfield  {title} {\bibinfo {title} {Disentangling orbital and valley hall effects in bilayers of transition metal dichalcogenides},\ }\href {https://doi.org/10.1103/PhysRevLett.126.056601} {\bibfield  {journal} {\bibinfo  {journal} {Phys. Rev. Lett.}\ }\textbf {\bibinfo {volume} {126}},\ \bibinfo {pages} {056601} (\bibinfo {year} {2021})}\BibitemShut {NoStop}%
\bibitem [{\citenamefont {Go}\ \emph {et~al.}(2018)\citenamefont {Go}, \citenamefont {Jo}, \citenamefont {Kim},\ and\ \citenamefont {Lee}}]{Go2018}%
  \BibitemOpen
  \bibfield  {author} {\bibinfo {author} {\bibfnamefont {D.}~\bibnamefont {Go}}, \bibinfo {author} {\bibfnamefont {D.}~\bibnamefont {Jo}}, \bibinfo {author} {\bibfnamefont {C.}~\bibnamefont {Kim}},\ and\ \bibinfo {author} {\bibfnamefont {H.-W.}\ \bibnamefont {Lee}},\ }\bibfield  {title} {\bibinfo {title} {Intrinsic spin and orbital hall effects from orbital texture},\ }\href {https://doi.org/10.1103/PhysRevLett.121.086602} {\bibfield  {journal} {\bibinfo  {journal} {Phys. Rev. Lett.}\ }\textbf {\bibinfo {volume} {121}},\ \bibinfo {pages} {086602} (\bibinfo {year} {2018})}\BibitemShut {NoStop}%
\bibitem [{\citenamefont {Abr\~ao}\ \emph {et~al.}(2025)\citenamefont {Abr\~ao}, \citenamefont {Santos}, \citenamefont {Costa}, \citenamefont {Santos}, \citenamefont {Mendes},\ and\ \citenamefont {Azevedo}}]{Abrao2025}%
  \BibitemOpen
  \bibfield  {author} {\bibinfo {author} {\bibfnamefont {J.~E.}\ \bibnamefont {Abr\~ao}}, \bibinfo {author} {\bibfnamefont {E.}~\bibnamefont {Santos}}, \bibinfo {author} {\bibfnamefont {J.~L.}\ \bibnamefont {Costa}}, \bibinfo {author} {\bibfnamefont {J.~G.~S.}\ \bibnamefont {Santos}}, \bibinfo {author} {\bibfnamefont {J.~B.~S.}\ \bibnamefont {Mendes}},\ and\ \bibinfo {author} {\bibfnamefont {A.}~\bibnamefont {Azevedo}},\ }\bibfield  {title} {\bibinfo {title} {Anomalous spin and orbital hall phenomena in antiferromagnetic systems},\ }\href {https://doi.org/10.1103/PhysRevLett.134.026702} {\bibfield  {journal} {\bibinfo  {journal} {Phys. Rev. Lett.}\ }\textbf {\bibinfo {volume} {134}},\ \bibinfo {pages} {026702} (\bibinfo {year} {2025})}\BibitemShut {NoStop}%
\bibitem [{\citenamefont {Phong}\ \emph {et~al.}(2019)\citenamefont {Phong}, \citenamefont {Addison}, \citenamefont {Ahn}, \citenamefont {Min}, \citenamefont {Agarwal},\ and\ \citenamefont {Mele}}]{Mele2019}%
  \BibitemOpen
  \bibfield  {author} {\bibinfo {author} {\bibfnamefont {V.~o.~T.}\ \bibnamefont {Phong}}, \bibinfo {author} {\bibfnamefont {Z.}~\bibnamefont {Addison}}, \bibinfo {author} {\bibfnamefont {S.}~\bibnamefont {Ahn}}, \bibinfo {author} {\bibfnamefont {H.}~\bibnamefont {Min}}, \bibinfo {author} {\bibfnamefont {R.}~\bibnamefont {Agarwal}},\ and\ \bibinfo {author} {\bibfnamefont {E.~J.}\ \bibnamefont {Mele}},\ }\bibfield  {title} {\bibinfo {title} {Optically controlled orbitronics on a triangular lattice},\ }\href {https://doi.org/10.1103/PhysRevLett.123.236403} {\bibfield  {journal} {\bibinfo  {journal} {Phys. Rev. Lett.}\ }\textbf {\bibinfo {volume} {123}},\ \bibinfo {pages} {236403} (\bibinfo {year} {2019})}\BibitemShut {NoStop}%
\bibitem [{\citenamefont {Bhowal}\ and\ \citenamefont {Vignale}(2021)}]{Bhowal2021}%
  \BibitemOpen
  \bibfield  {author} {\bibinfo {author} {\bibfnamefont {S.}~\bibnamefont {Bhowal}}\ and\ \bibinfo {author} {\bibfnamefont {G.}~\bibnamefont {Vignale}},\ }\bibfield  {title} {\bibinfo {title} {Orbital hall effect as an alternative to valley hall effect in gapped graphene},\ }\href {https://doi.org/10.1103/PhysRevB.103.195309} {\bibfield  {journal} {\bibinfo  {journal} {Phys. Rev. B}\ }\textbf {\bibinfo {volume} {103}},\ \bibinfo {pages} {195309} (\bibinfo {year} {2021})}\BibitemShut {NoStop}%
\bibitem [{\citenamefont {Johansson}(2024)}]{Johansson2024}%
  \BibitemOpen
  \bibfield  {author} {\bibinfo {author} {\bibfnamefont {A.}~\bibnamefont {Johansson}},\ }\bibfield  {title} {\bibinfo {title} {Theory of spin and orbital edelstein effects},\ }\href {https://doi.org/10.1088/1361-648X/ad5e2b} {\bibfield  {journal} {\bibinfo  {journal} {Journal of Physics: Condensed Matter}\ }\textbf {\bibinfo {volume} {36}},\ \bibinfo {pages} {423002} (\bibinfo {year} {2024})}\BibitemShut {NoStop}%
\bibitem [{\citenamefont {Salemi}\ \emph {et~al.}(2019)\citenamefont {Salemi}, \citenamefont {Berritta}, \citenamefont {Nandy},\ and\ \citenamefont {Oppeneer}}]{Salemi2019}%
  \BibitemOpen
  \bibfield  {author} {\bibinfo {author} {\bibfnamefont {L.}~\bibnamefont {Salemi}}, \bibinfo {author} {\bibfnamefont {M.}~\bibnamefont {Berritta}}, \bibinfo {author} {\bibfnamefont {A.~K.}\ \bibnamefont {Nandy}},\ and\ \bibinfo {author} {\bibfnamefont {P.~M.}\ \bibnamefont {Oppeneer}},\ }\bibfield  {title} {\bibinfo {title} {Orbitally dominated rashba-edelstein effect in noncentrosymmetric antiferromagnets},\ }\bibfield  {journal} {\bibinfo  {journal} {Nature Communications}\ }\textbf {\bibinfo {volume} {10}},\ \href {https://doi.org/10.1038/s41467-019-13367-z} {10.1038/s41467-019-13367-z} (\bibinfo {year} {2019})\BibitemShut {NoStop}%
\bibitem [{\citenamefont {Nikolaev}\ \emph {et~al.}(2024)\citenamefont {Nikolaev}, \citenamefont {Chshiev}, \citenamefont {Ibrahim}, \citenamefont {Krishnia}, \citenamefont {Sebe}, \citenamefont {George}, \citenamefont {Cros}, \citenamefont {Jaffrès},\ and\ \citenamefont {Fert}}]{Nikolaev2024}%
  \BibitemOpen
  \bibfield  {author} {\bibinfo {author} {\bibfnamefont {S.~A.}\ \bibnamefont {Nikolaev}}, \bibinfo {author} {\bibfnamefont {M.}~\bibnamefont {Chshiev}}, \bibinfo {author} {\bibfnamefont {F.}~\bibnamefont {Ibrahim}}, \bibinfo {author} {\bibfnamefont {S.}~\bibnamefont {Krishnia}}, \bibinfo {author} {\bibfnamefont {N.}~\bibnamefont {Sebe}}, \bibinfo {author} {\bibfnamefont {J.-M.}\ \bibnamefont {George}}, \bibinfo {author} {\bibfnamefont {V.}~\bibnamefont {Cros}}, \bibinfo {author} {\bibfnamefont {H.}~\bibnamefont {Jaffrès}},\ and\ \bibinfo {author} {\bibfnamefont {A.}~\bibnamefont {Fert}},\ }\bibfield  {title} {\bibinfo {title} {Large chiral orbital texture and orbital edelstein effect in co/al heterostructure},\ }\href {https://doi.org/10.1021/acs.nanolett.4c01607} {\bibfield  {journal} {\bibinfo  {journal} {Nano Letters}\ }\textbf {\bibinfo {volume} {24}},\ \bibinfo {pages} {13465} (\bibinfo {year} {2024})},\ \bibinfo {note} {pMID: 39433297},\ \Eprint
  {https://arxiv.org/abs/https://doi.org/10.1021/acs.nanolett.4c01607} {https://doi.org/10.1021/acs.nanolett.4c01607} \BibitemShut {NoStop}%
\bibitem [{\citenamefont {Xiao}\ \emph {et~al.}(2026)\citenamefont {Xiao}, \citenamefont {Zhao}, \citenamefont {Baek}, \citenamefont {Lee}, \citenamefont {Zheng}, \citenamefont {Shi}, \citenamefont {Xie}, \citenamefont {Zhang}, \citenamefont {Jia}, \citenamefont {Yang}, \citenamefont {Yu}, \citenamefont {Song}, \citenamefont {Lee},\ and\ \citenamefont {Chen}}]{Xiao2026}%
  \BibitemOpen
  \bibfield  {author} {\bibinfo {author} {\bibfnamefont {R.}~\bibnamefont {Xiao}}, \bibinfo {author} {\bibfnamefont {T.}~\bibnamefont {Zhao}}, \bibinfo {author} {\bibfnamefont {I.}~\bibnamefont {Baek}}, \bibinfo {author} {\bibfnamefont {H.}~\bibnamefont {Lee}}, \bibinfo {author} {\bibfnamefont {Z.}~\bibnamefont {Zheng}}, \bibinfo {author} {\bibfnamefont {S.}~\bibnamefont {Shi}}, \bibinfo {author} {\bibfnamefont {Z.}~\bibnamefont {Xie}}, \bibinfo {author} {\bibfnamefont {Q.}~\bibnamefont {Zhang}}, \bibinfo {author} {\bibfnamefont {L.}~\bibnamefont {Jia}}, \bibinfo {author} {\bibfnamefont {P.}~\bibnamefont {Yang}}, \bibinfo {author} {\bibfnamefont {X.}~\bibnamefont {Yu}}, \bibinfo {author} {\bibfnamefont {D.}~\bibnamefont {Song}}, \bibinfo {author} {\bibfnamefont {H.-W.}\ \bibnamefont {Lee}},\ and\ \bibinfo {author} {\bibfnamefont {J.}~\bibnamefont {Chen}},\ }\bibfield  {title} {\bibinfo {title} {Crystal symmetry-dependent orbital rashba edelstein effect in epitaxial cuo thin film},\ }\bibfield  {journal}
  {\bibinfo  {journal} {Nature Communications}\ }\textbf {\bibinfo {volume} {17}},\ \href {https://doi.org/10.1038/s41467-026-71018-6} {10.1038/s41467-026-71018-6} (\bibinfo {year} {2026})\BibitemShut {NoStop}%
\bibitem [{\citenamefont {Lee}\ \emph {et~al.}(2021)\citenamefont {Lee}, \citenamefont {Go}, \citenamefont {Park}, \citenamefont {Jeong}, \citenamefont {Ko}, \citenamefont {Yun}, \citenamefont {Jo}, \citenamefont {Lee}, \citenamefont {Go}, \citenamefont {Oh}, \citenamefont {Kim}, \citenamefont {Park}, \citenamefont {Min}, \citenamefont {Koo}, \citenamefont {Lee}, \citenamefont {Lee},\ and\ \citenamefont {Lee}}]{Lee2021}%
  \BibitemOpen
  \bibfield  {author} {\bibinfo {author} {\bibfnamefont {D.}~\bibnamefont {Lee}}, \bibinfo {author} {\bibfnamefont {D.}~\bibnamefont {Go}}, \bibinfo {author} {\bibfnamefont {H.-J.}\ \bibnamefont {Park}}, \bibinfo {author} {\bibfnamefont {W.}~\bibnamefont {Jeong}}, \bibinfo {author} {\bibfnamefont {H.-W.}\ \bibnamefont {Ko}}, \bibinfo {author} {\bibfnamefont {D.}~\bibnamefont {Yun}}, \bibinfo {author} {\bibfnamefont {D.}~\bibnamefont {Jo}}, \bibinfo {author} {\bibfnamefont {S.}~\bibnamefont {Lee}}, \bibinfo {author} {\bibfnamefont {G.}~\bibnamefont {Go}}, \bibinfo {author} {\bibfnamefont {J.~H.}\ \bibnamefont {Oh}}, \bibinfo {author} {\bibfnamefont {K.-J.}\ \bibnamefont {Kim}}, \bibinfo {author} {\bibfnamefont {B.-G.}\ \bibnamefont {Park}}, \bibinfo {author} {\bibfnamefont {B.-C.}\ \bibnamefont {Min}}, \bibinfo {author} {\bibfnamefont {H.~C.}\ \bibnamefont {Koo}}, \bibinfo {author} {\bibfnamefont {H.-W.}\ \bibnamefont {Lee}}, \bibinfo {author} {\bibfnamefont {O.}~\bibnamefont {Lee}},\ and\ \bibinfo {author}
  {\bibfnamefont {K.-J.}\ \bibnamefont {Lee}},\ }\bibfield  {title} {\bibinfo {title} {Orbital torque in magnetic bilayers},\ }\href {https://doi.org/10.1038/s41467-021-26650-9} {\bibfield  {journal} {\bibinfo  {journal} {Nature Communications}\ }\textbf {\bibinfo {volume} {12}},\ \bibinfo {pages} {6710} (\bibinfo {year} {2021})}\BibitemShut {NoStop}%
\bibitem [{\citenamefont {Yang}\ \emph {et~al.}(2024)\citenamefont {Yang}, \citenamefont {Wang}, \citenamefont {Chen}, \citenamefont {Zhang}, \citenamefont {Pan}, \citenamefont {Hu}, \citenamefont {Wang}, \citenamefont {Yue}, \citenamefont {Chen}, \citenamefont {Jiang}, \citenamefont {Zhu}, \citenamefont {Qiu}, \citenamefont {Yao}, \citenamefont {Li}, \citenamefont {Wang},\ and\ \citenamefont {Jiang}}]{Yang2024}%
  \BibitemOpen
  \bibfield  {author} {\bibinfo {author} {\bibfnamefont {Y.}~\bibnamefont {Yang}}, \bibinfo {author} {\bibfnamefont {P.}~\bibnamefont {Wang}}, \bibinfo {author} {\bibfnamefont {J.}~\bibnamefont {Chen}}, \bibinfo {author} {\bibfnamefont {D.}~\bibnamefont {Zhang}}, \bibinfo {author} {\bibfnamefont {C.}~\bibnamefont {Pan}}, \bibinfo {author} {\bibfnamefont {S.}~\bibnamefont {Hu}}, \bibinfo {author} {\bibfnamefont {T.}~\bibnamefont {Wang}}, \bibinfo {author} {\bibfnamefont {W.}~\bibnamefont {Yue}}, \bibinfo {author} {\bibfnamefont {C.}~\bibnamefont {Chen}}, \bibinfo {author} {\bibfnamefont {W.}~\bibnamefont {Jiang}}, \bibinfo {author} {\bibfnamefont {L.}~\bibnamefont {Zhu}}, \bibinfo {author} {\bibfnamefont {X.}~\bibnamefont {Qiu}}, \bibinfo {author} {\bibfnamefont {Y.}~\bibnamefont {Yao}}, \bibinfo {author} {\bibfnamefont {Y.}~\bibnamefont {Li}}, \bibinfo {author} {\bibfnamefont {W.}~\bibnamefont {Wang}},\ and\ \bibinfo {author} {\bibfnamefont {Y.}~\bibnamefont {Jiang}},\ }\bibfield  {title} {\bibinfo {title}
  {Orbital torque switching in perpendicularly magnetized materials},\ }\bibfield  {journal} {\bibinfo  {journal} {Nature Communications}\ }\textbf {\bibinfo {volume} {15}},\ \href {https://doi.org/10.1038/s41467-024-52824-2} {10.1038/s41467-024-52824-2} (\bibinfo {year} {2024})\BibitemShut {NoStop}%
\bibitem [{\citenamefont {Go}\ and\ \citenamefont {Lee}(2020)}]{Go2020}%
  \BibitemOpen
  \bibfield  {author} {\bibinfo {author} {\bibfnamefont {D.}~\bibnamefont {Go}}\ and\ \bibinfo {author} {\bibfnamefont {H.-W.}\ \bibnamefont {Lee}},\ }\bibfield  {title} {\bibinfo {title} {Orbital torque: Torque generation by orbital current injection},\ }\href {https://doi.org/10.1103/PhysRevResearch.2.013177} {\bibfield  {journal} {\bibinfo  {journal} {Phys. Rev. Res.}\ }\textbf {\bibinfo {volume} {2}},\ \bibinfo {pages} {013177} (\bibinfo {year} {2020})}\BibitemShut {NoStop}%
\bibitem [{\citenamefont {Go}\ \emph {et~al.}(2025)\citenamefont {Go}, \citenamefont {Ando}, \citenamefont {Pezo}, \citenamefont {Bl\"ugel}, \citenamefont {Manchon},\ and\ \citenamefont {Mokrousov}}]{Go2023}%
  \BibitemOpen
  \bibfield  {author} {\bibinfo {author} {\bibfnamefont {D.}~\bibnamefont {Go}}, \bibinfo {author} {\bibfnamefont {K.}~\bibnamefont {Ando}}, \bibinfo {author} {\bibfnamefont {A.}~\bibnamefont {Pezo}}, \bibinfo {author} {\bibfnamefont {S.}~\bibnamefont {Bl\"ugel}}, \bibinfo {author} {\bibfnamefont {A.}~\bibnamefont {Manchon}},\ and\ \bibinfo {author} {\bibfnamefont {Y.}~\bibnamefont {Mokrousov}},\ }\bibfield  {title} {\bibinfo {title} {Orbital pumping by magnetization dynamics in ferromagnets},\ }\href {https://doi.org/10.1103/PhysRevB.111.L140409} {\bibfield  {journal} {\bibinfo  {journal} {Phys. Rev. B}\ }\textbf {\bibinfo {volume} {111}},\ \bibinfo {pages} {L140409} (\bibinfo {year} {2025})}\BibitemShut {NoStop}%
\bibitem [{\citenamefont {Santos}\ \emph {et~al.}(2023)\citenamefont {Santos}, \citenamefont {Abr\~ao}, \citenamefont {Go}, \citenamefont {de~Assis}, \citenamefont {Mokrousov}, \citenamefont {Mendes},\ and\ \citenamefont {Azevedo}}]{Santos2023}%
  \BibitemOpen
  \bibfield  {author} {\bibinfo {author} {\bibfnamefont {E.}~\bibnamefont {Santos}}, \bibinfo {author} {\bibfnamefont {J.}~\bibnamefont {Abr\~ao}}, \bibinfo {author} {\bibfnamefont {D.}~\bibnamefont {Go}}, \bibinfo {author} {\bibfnamefont {L.}~\bibnamefont {de~Assis}}, \bibinfo {author} {\bibfnamefont {Y.}~\bibnamefont {Mokrousov}}, \bibinfo {author} {\bibfnamefont {J.}~\bibnamefont {Mendes}},\ and\ \bibinfo {author} {\bibfnamefont {A.}~\bibnamefont {Azevedo}},\ }\bibfield  {title} {\bibinfo {title} {Inverse orbital torque via spin-orbital intertwined states},\ }\href {https://doi.org/10.1103/PhysRevApplied.19.014069} {\bibfield  {journal} {\bibinfo  {journal} {Phys. Rev. Appl.}\ }\textbf {\bibinfo {volume} {19}},\ \bibinfo {pages} {014069} (\bibinfo {year} {2023})}\BibitemShut {NoStop}%
\bibitem [{\citenamefont {Ding}\ \emph {et~al.}(2025)\citenamefont {Ding}, \citenamefont {Noël}, \citenamefont {Krishnaswamy}, \citenamefont {Davitti}, \citenamefont {Sala}, \citenamefont {Fantauzzi}, \citenamefont {Rossi},\ and\ \citenamefont {Gambardella}}]{Ding2025CoO}%
  \BibitemOpen
  \bibfield  {author} {\bibinfo {author} {\bibfnamefont {S.}~\bibnamefont {Ding}}, \bibinfo {author} {\bibfnamefont {P.}~\bibnamefont {Noël}}, \bibinfo {author} {\bibfnamefont {G.~K.}\ \bibnamefont {Krishnaswamy}}, \bibinfo {author} {\bibfnamefont {N.}~\bibnamefont {Davitti}}, \bibinfo {author} {\bibfnamefont {G.}~\bibnamefont {Sala}}, \bibinfo {author} {\bibfnamefont {M.}~\bibnamefont {Fantauzzi}}, \bibinfo {author} {\bibfnamefont {A.}~\bibnamefont {Rossi}},\ and\ \bibinfo {author} {\bibfnamefont {P.}~\bibnamefont {Gambardella}},\ }\bibfield  {title} {\bibinfo {title} {Generation, transmission, and conversion of orbital torque by an antiferromagnetic insulator},\ }\bibfield  {journal} {\bibinfo  {journal} {Nature Communications}\ }\textbf {\bibinfo {volume} {16}},\ \href {https://doi.org/10.1038/s41467-025-64273-6} {10.1038/s41467-025-64273-6} (\bibinfo {year} {2025})\BibitemShut {NoStop}%
\bibitem [{\citenamefont {Schmitt}\ \emph {et~al.}(2026)\citenamefont {Schmitt}, \citenamefont {Krishnia}, \citenamefont {Zeer}, \citenamefont {Galíndez-Ruales}, \citenamefont {Loyal}, \citenamefont {Köhler}, \citenamefont {Micus}, \citenamefont {Kikkawa}, \citenamefont {Arisawa}, \citenamefont {Denneulin}, \citenamefont {Kovács}, \citenamefont {Xu}, \citenamefont {Tran}, \citenamefont {Kronast}, \citenamefont {Go}, \citenamefont {Pourovskii}, \citenamefont {Dunin-Borkowski}, \citenamefont {Kuschel}, \citenamefont {Ležaić}, \citenamefont {Sinova}, \citenamefont {Saitoh}, \citenamefont {Jakob}, \citenamefont {Gomonay}, \citenamefont {Mokrousov},\ and\ \citenamefont {Kläui}}]{Schmitt2026CoO}%
  \BibitemOpen
  \bibfield  {author} {\bibinfo {author} {\bibfnamefont {C.}~\bibnamefont {Schmitt}}, \bibinfo {author} {\bibfnamefont {S.}~\bibnamefont {Krishnia}}, \bibinfo {author} {\bibfnamefont {M.}~\bibnamefont {Zeer}}, \bibinfo {author} {\bibfnamefont {E.}~\bibnamefont {Galíndez-Ruales}}, \bibinfo {author} {\bibfnamefont {M.}~\bibnamefont {Loyal}}, \bibinfo {author} {\bibfnamefont {J.}~\bibnamefont {Köhler}}, \bibinfo {author} {\bibfnamefont {L.}~\bibnamefont {Micus}}, \bibinfo {author} {\bibfnamefont {T.}~\bibnamefont {Kikkawa}}, \bibinfo {author} {\bibfnamefont {H.}~\bibnamefont {Arisawa}}, \bibinfo {author} {\bibfnamefont {T.}~\bibnamefont {Denneulin}}, \bibinfo {author} {\bibfnamefont {A.}~\bibnamefont {Kovács}}, \bibinfo {author} {\bibfnamefont {R.}~\bibnamefont {Xu}}, \bibinfo {author} {\bibfnamefont {D.}~\bibnamefont {Tran}}, \bibinfo {author} {\bibfnamefont {F.}~\bibnamefont {Kronast}}, \bibinfo {author} {\bibfnamefont {D.}~\bibnamefont {Go}}, \bibinfo {author} {\bibfnamefont {L.~V.}\ \bibnamefont
  {Pourovskii}}, \bibinfo {author} {\bibfnamefont {R.~E.}\ \bibnamefont {Dunin-Borkowski}}, \bibinfo {author} {\bibfnamefont {T.}~\bibnamefont {Kuschel}}, \bibinfo {author} {\bibfnamefont {M.}~\bibnamefont {Ležaić}}, \bibinfo {author} {\bibfnamefont {J.}~\bibnamefont {Sinova}}, \bibinfo {author} {\bibfnamefont {E.}~\bibnamefont {Saitoh}}, \bibinfo {author} {\bibfnamefont {G.}~\bibnamefont {Jakob}}, \bibinfo {author} {\bibfnamefont {O.}~\bibnamefont {Gomonay}}, \bibinfo {author} {\bibfnamefont {Y.}~\bibnamefont {Mokrousov}},\ and\ \bibinfo {author} {\bibfnamefont {M.}~\bibnamefont {Kläui}},\ }\bibfield  {title} {\bibinfo {title} {Orbital magnetoresistance in the antiferromagnet coo driven by dynamic orbital angular momentum},\ }\href {https://doi.org/10.1126/science.adw1808} {\bibfield  {journal} {\bibinfo  {journal} {Science}\ }\textbf {\bibinfo {volume} {393}},\ \bibinfo {pages} {76} (\bibinfo {year} {2026})},\ \Eprint {https://arxiv.org/abs/https://www.science.org/doi/pdf/10.1126/science.adw1808}
  {https://www.science.org/doi/pdf/10.1126/science.adw1808} \BibitemShut {NoStop}%
\bibitem [{\citenamefont {Šmejkal}\ \emph {et~al.}(2020)\citenamefont {Šmejkal}, \citenamefont {González-Hernández}, \citenamefont {Jungwirth},\ and\ \citenamefont {Sinova}}]{Smejkal2020}%
  \BibitemOpen
  \bibfield  {author} {\bibinfo {author} {\bibfnamefont {L.}~\bibnamefont {Šmejkal}}, \bibinfo {author} {\bibfnamefont {R.}~\bibnamefont {González-Hernández}}, \bibinfo {author} {\bibfnamefont {T.}~\bibnamefont {Jungwirth}},\ and\ \bibinfo {author} {\bibfnamefont {J.}~\bibnamefont {Sinova}},\ }\bibfield  {title} {\bibinfo {title} {Crystal time-reversal symmetry breaking and spontaneous hall effect in collinear antiferromagnets},\ }\href {https://doi.org/10.1126/sciadv.aaz8809} {\bibfield  {journal} {\bibinfo  {journal} {Science Advances}\ }\textbf {\bibinfo {volume} {6}},\ \bibinfo {pages} {eaaz8809} (\bibinfo {year} {2020})},\ \Eprint {https://arxiv.org/abs/https://www.science.org/doi/pdf/10.1126/sciadv.aaz8809} {https://www.science.org/doi/pdf/10.1126/sciadv.aaz8809} \BibitemShut {NoStop}%
\bibitem [{\citenamefont {Hayami}\ \emph {et~al.}(2019)\citenamefont {Hayami}, \citenamefont {Yanagi},\ and\ \citenamefont {Kusunose}}]{Hayami2019}%
  \BibitemOpen
  \bibfield  {author} {\bibinfo {author} {\bibfnamefont {S.}~\bibnamefont {Hayami}}, \bibinfo {author} {\bibfnamefont {Y.}~\bibnamefont {Yanagi}},\ and\ \bibinfo {author} {\bibfnamefont {H.}~\bibnamefont {Kusunose}},\ }\bibfield  {title} {\bibinfo {title} {Momentum-dependent spin splitting by collinear antiferromagnetic ordering},\ }\bibfield  {journal} {\bibinfo  {journal} {Journal of the Physical Society of Japan}\ }\textbf {\bibinfo {volume} {88}},\ \href {https://doi.org/10.7566/jpsj.88.123702} {10.7566/jpsj.88.123702} (\bibinfo {year} {2019})\BibitemShut {NoStop}%
\bibitem [{\citenamefont {Yuan}\ \emph {et~al.}(2020)\citenamefont {Yuan}, \citenamefont {Wang}, \citenamefont {Luo}, \citenamefont {Rashba},\ and\ \citenamefont {Zunger}}]{Yuan2020}%
  \BibitemOpen
  \bibfield  {author} {\bibinfo {author} {\bibfnamefont {L.-D.}\ \bibnamefont {Yuan}}, \bibinfo {author} {\bibfnamefont {Z.}~\bibnamefont {Wang}}, \bibinfo {author} {\bibfnamefont {J.-W.}\ \bibnamefont {Luo}}, \bibinfo {author} {\bibfnamefont {E.~I.}\ \bibnamefont {Rashba}},\ and\ \bibinfo {author} {\bibfnamefont {A.}~\bibnamefont {Zunger}},\ }\bibfield  {title} {\bibinfo {title} {Giant momentum-dependent spin splitting in centrosymmetric low-$z$ antiferromagnets},\ }\href {https://doi.org/10.1103/PhysRevB.102.014422} {\bibfield  {journal} {\bibinfo  {journal} {Phys. Rev. B}\ }\textbf {\bibinfo {volume} {102}},\ \bibinfo {pages} {014422} (\bibinfo {year} {2020})}\BibitemShut {NoStop}%
\bibitem [{\citenamefont {\v{S}mejkal}\ \emph {et~al.}(2022)\citenamefont {\v{S}mejkal}, \citenamefont {Sinova},\ and\ \citenamefont {Jungwirth}}]{Smejkal2022}%
  \BibitemOpen
  \bibfield  {author} {\bibinfo {author} {\bibfnamefont {L.}~\bibnamefont {\v{S}mejkal}}, \bibinfo {author} {\bibfnamefont {J.}~\bibnamefont {Sinova}},\ and\ \bibinfo {author} {\bibfnamefont {T.}~\bibnamefont {Jungwirth}},\ }\bibfield  {title} {\bibinfo {title} {Emerging research landscape of altermagnetism},\ }\href {https://doi.org/10.1103/PhysRevX.12.040501} {\bibfield  {journal} {\bibinfo  {journal} {Physical Review X}\ }\textbf {\bibinfo {volume} {12}},\ \bibinfo {pages} {040501} (\bibinfo {year} {2022})}\BibitemShut {NoStop}%
\bibitem [{\citenamefont {Li}\ and\ \citenamefont {Sukhachov}(2026)}]{Li2026}%
  \BibitemOpen
  \bibfield  {author} {\bibinfo {author} {\bibfnamefont {Y.}~\bibnamefont {Li}}\ and\ \bibinfo {author} {\bibfnamefont {P.}~\bibnamefont {Sukhachov}},\ }\href {https://arxiv.org/abs/2604.18695} {\bibinfo {title} {$p$-wave orbital magnetism}} (\bibinfo {year} {2026}),\ \Eprint {https://arxiv.org/abs/2604.18695} {arXiv:2604.18695 [cond-mat.mes-hall]} \BibitemShut {NoStop}%
\bibitem [{\citenamefont {Pan}\ \emph {et~al.}(2026)\citenamefont {Pan}, \citenamefont {Liu},\ and\ \citenamefont {Huang}}]{Pan2026}%
  \BibitemOpen
  \bibfield  {author} {\bibinfo {author} {\bibfnamefont {M.}~\bibnamefont {Pan}}, \bibinfo {author} {\bibfnamefont {F.}~\bibnamefont {Liu}},\ and\ \bibinfo {author} {\bibfnamefont {H.}~\bibnamefont {Huang}},\ }\bibfield  {title} {\bibinfo {title} {Orbital altermagnetism in two dimensions},\ }\href {https://doi.org/10.1103/l8fc-dp36} {\bibfield  {journal} {\bibinfo  {journal} {Phys. Rev. Lett.}\ }\textbf {\bibinfo {volume} {137}},\ \bibinfo {pages} {016702} (\bibinfo {year} {2026})}\BibitemShut {NoStop}%
\bibitem [{\citenamefont {Xiao}\ \emph {et~al.}(2005)\citenamefont {Xiao}, \citenamefont {Shi},\ and\ \citenamefont {Niu}}]{Xiao2005}%
  \BibitemOpen
  \bibfield  {author} {\bibinfo {author} {\bibfnamefont {D.}~\bibnamefont {Xiao}}, \bibinfo {author} {\bibfnamefont {J.}~\bibnamefont {Shi}},\ and\ \bibinfo {author} {\bibfnamefont {Q.}~\bibnamefont {Niu}},\ }\bibfield  {title} {\bibinfo {title} {Berry phase correction to electron density of states in solids},\ }\href {https://doi.org/10.1103/PhysRevLett.95.137204} {\bibfield  {journal} {\bibinfo  {journal} {Phys. Rev. Lett.}\ }\textbf {\bibinfo {volume} {95}},\ \bibinfo {pages} {137204} (\bibinfo {year} {2005})}\BibitemShut {NoStop}%
\bibitem [{\citenamefont {Thonhauser}\ \emph {et~al.}(2005)\citenamefont {Thonhauser}, \citenamefont {Ceresoli}, \citenamefont {Vanderbilt},\ and\ \citenamefont {Resta}}]{Thonhauser2005}%
  \BibitemOpen
  \bibfield  {author} {\bibinfo {author} {\bibfnamefont {T.}~\bibnamefont {Thonhauser}}, \bibinfo {author} {\bibfnamefont {D.}~\bibnamefont {Ceresoli}}, \bibinfo {author} {\bibfnamefont {D.}~\bibnamefont {Vanderbilt}},\ and\ \bibinfo {author} {\bibfnamefont {R.}~\bibnamefont {Resta}},\ }\bibfield  {title} {\bibinfo {title} {Orbital magnetization in periodic insulators},\ }\href {https://doi.org/10.1103/PhysRevLett.95.137205} {\bibfield  {journal} {\bibinfo  {journal} {Phys. Rev. Lett.}\ }\textbf {\bibinfo {volume} {95}},\ \bibinfo {pages} {137205} (\bibinfo {year} {2005})}\BibitemShut {NoStop}%
\bibitem [{\citenamefont {Ceresoli}\ \emph {et~al.}(2006)\citenamefont {Ceresoli}, \citenamefont {Thonhauser}, \citenamefont {Vanderbilt},\ and\ \citenamefont {Resta}}]{Ceresoli2006}%
  \BibitemOpen
  \bibfield  {author} {\bibinfo {author} {\bibfnamefont {D.}~\bibnamefont {Ceresoli}}, \bibinfo {author} {\bibfnamefont {T.}~\bibnamefont {Thonhauser}}, \bibinfo {author} {\bibfnamefont {D.}~\bibnamefont {Vanderbilt}},\ and\ \bibinfo {author} {\bibfnamefont {R.}~\bibnamefont {Resta}},\ }\bibfield  {title} {\bibinfo {title} {Orbital magnetization in crystalline solids: Multi-band insulators, chern insulators, and metals},\ }\href {https://doi.org/10.1103/PhysRevB.74.024408} {\bibfield  {journal} {\bibinfo  {journal} {Phys. Rev. B}\ }\textbf {\bibinfo {volume} {74}},\ \bibinfo {pages} {024408} (\bibinfo {year} {2006})}\BibitemShut {NoStop}%
\bibitem [{\citenamefont {Souza}\ and\ \citenamefont {Vanderbilt}(2008)}]{Souza2008}%
  \BibitemOpen
  \bibfield  {author} {\bibinfo {author} {\bibfnamefont {I.}~\bibnamefont {Souza}}\ and\ \bibinfo {author} {\bibfnamefont {D.}~\bibnamefont {Vanderbilt}},\ }\bibfield  {title} {\bibinfo {title} {Dichroic $f$-sum rule and the orbital magnetization of crystals},\ }\href {https://doi.org/10.1103/PhysRevB.77.054438} {\bibfield  {journal} {\bibinfo  {journal} {Phys. Rev. B}\ }\textbf {\bibinfo {volume} {77}},\ \bibinfo {pages} {054438} (\bibinfo {year} {2008})}\BibitemShut {NoStop}%
\bibitem [{\citenamefont {Haldane}(1988)}]{Haldane1988}%
  \BibitemOpen
  \bibfield  {author} {\bibinfo {author} {\bibfnamefont {F.~D.~M.}\ \bibnamefont {Haldane}},\ }\bibfield  {title} {\bibinfo {title} {Model for a quantum hall effect without landau levels: Condensed-matter realization of the "parity anomaly"},\ }\href {https://doi.org/10.1103/PhysRevLett.61.2015} {\bibfield  {journal} {\bibinfo  {journal} {Phys. Rev. Lett.}\ }\textbf {\bibinfo {volume} {61}},\ \bibinfo {pages} {2015} (\bibinfo {year} {1988})}\BibitemShut {NoStop}%
\bibitem [{\citenamefont {Colom\'es}\ and\ \citenamefont {Franz}(2018)}]{Colomes2018}%
  \BibitemOpen
  \bibfield  {author} {\bibinfo {author} {\bibfnamefont {E.}~\bibnamefont {Colom\'es}}\ and\ \bibinfo {author} {\bibfnamefont {M.}~\bibnamefont {Franz}},\ }\bibfield  {title} {\bibinfo {title} {Antichiral edge states in a modified haldane nanoribbon},\ }\href {https://doi.org/10.1103/PhysRevLett.120.086603} {\bibfield  {journal} {\bibinfo  {journal} {Phys. Rev. Lett.}\ }\textbf {\bibinfo {volume} {120}},\ \bibinfo {pages} {086603} (\bibinfo {year} {2018})}\BibitemShut {NoStop}%
\bibitem [{\citenamefont {Bianco}\ and\ \citenamefont {Resta}(2013)}]{Bianco2013}%
  \BibitemOpen
  \bibfield  {author} {\bibinfo {author} {\bibfnamefont {R.}~\bibnamefont {Bianco}}\ and\ \bibinfo {author} {\bibfnamefont {R.}~\bibnamefont {Resta}},\ }\bibfield  {title} {\bibinfo {title} {Orbital magnetization as a local property},\ }\href {https://doi.org/10.1103/PhysRevLett.110.087202} {\bibfield  {journal} {\bibinfo  {journal} {Phys. Rev. Lett.}\ }\textbf {\bibinfo {volume} {110}},\ \bibinfo {pages} {087202} (\bibinfo {year} {2013})}\BibitemShut {NoStop}%
\bibitem [{\citenamefont {Vidarte}\ \emph {et~al.}(2026)\citenamefont {Vidarte}, \citenamefont {Veiga}, \citenamefont {Lopes}, \citenamefont {Cardias}, \citenamefont {Ferreira}, \citenamefont {Cysne},\ and\ \citenamefont {Rappoport}}]{Vidarte2026}%
  \BibitemOpen
  \bibfield  {author} {\bibinfo {author} {\bibfnamefont {K.~J.~U.}\ \bibnamefont {Vidarte}}, \bibinfo {author} {\bibfnamefont {H.~P.}\ \bibnamefont {Veiga}}, \bibinfo {author} {\bibfnamefont {J.~a. M. V.~P.}\ \bibnamefont {Lopes}}, \bibinfo {author} {\bibfnamefont {R.}~\bibnamefont {Cardias}}, \bibinfo {author} {\bibfnamefont {A.}~\bibnamefont {Ferreira}}, \bibinfo {author} {\bibfnamefont {T.~P.}\ \bibnamefont {Cysne}},\ and\ \bibinfo {author} {\bibfnamefont {T.~G.}\ \bibnamefont {Rappoport}},\ }\bibfield  {title} {\bibinfo {title} {Real-space spectral approach to orbital magnetization},\ }\href {https://doi.org/10.1103/bhg8-c3rw} {\bibfield  {journal} {\bibinfo  {journal} {Phys. Rev. B}\ }\textbf {\bibinfo {volume} {113}},\ \bibinfo {pages} {224438} (\bibinfo {year} {2026})}\BibitemShut {NoStop}%
\bibitem [{\citenamefont {Vila}\ \emph {et~al.}(2019)\citenamefont {Vila}, \citenamefont {Hung}, \citenamefont {Roche},\ and\ \citenamefont {Saito}}]{Vila2019}%
  \BibitemOpen
  \bibfield  {author} {\bibinfo {author} {\bibfnamefont {M.}~\bibnamefont {Vila}}, \bibinfo {author} {\bibfnamefont {N.~T.}\ \bibnamefont {Hung}}, \bibinfo {author} {\bibfnamefont {S.}~\bibnamefont {Roche}},\ and\ \bibinfo {author} {\bibfnamefont {R.}~\bibnamefont {Saito}},\ }\bibfield  {title} {\bibinfo {title} {Tunable circular dichroism and valley polarization in the modified haldane model},\ }\href {https://doi.org/10.1103/PhysRevB.99.161404} {\bibfield  {journal} {\bibinfo  {journal} {Phys. Rev. B}\ }\textbf {\bibinfo {volume} {99}},\ \bibinfo {pages} {161404(R)} (\bibinfo {year} {2019})}\BibitemShut {NoStop}%
\bibitem [{\citenamefont {Zhou}\ \emph {et~al.}(2020)\citenamefont {Zhou}, \citenamefont {Liu}, \citenamefont {Yang}, \citenamefont {Hu}, \citenamefont {Ma}, \citenamefont {Xue}, \citenamefont {Wang}, \citenamefont {Deng},\ and\ \citenamefont {Zhang}}]{Zhou2020}%
  \BibitemOpen
  \bibfield  {author} {\bibinfo {author} {\bibfnamefont {P.}~\bibnamefont {Zhou}}, \bibinfo {author} {\bibfnamefont {G.-G.}\ \bibnamefont {Liu}}, \bibinfo {author} {\bibfnamefont {Y.}~\bibnamefont {Yang}}, \bibinfo {author} {\bibfnamefont {Y.-H.}\ \bibnamefont {Hu}}, \bibinfo {author} {\bibfnamefont {S.}~\bibnamefont {Ma}}, \bibinfo {author} {\bibfnamefont {H.}~\bibnamefont {Xue}}, \bibinfo {author} {\bibfnamefont {Q.}~\bibnamefont {Wang}}, \bibinfo {author} {\bibfnamefont {L.}~\bibnamefont {Deng}},\ and\ \bibinfo {author} {\bibfnamefont {B.}~\bibnamefont {Zhang}},\ }\bibfield  {title} {\bibinfo {title} {Observation of photonic antichiral edge states},\ }\href {https://doi.org/10.1103/PhysRevLett.125.263603} {\bibfield  {journal} {\bibinfo  {journal} {Phys. Rev. Lett.}\ }\textbf {\bibinfo {volume} {125}},\ \bibinfo {pages} {263603} (\bibinfo {year} {2020})}\BibitemShut {NoStop}%
\bibitem [{\citenamefont {Frank}\ \emph {et~al.}(2018)\citenamefont {Frank}, \citenamefont {H\"ogl}, \citenamefont {Gmitra}, \citenamefont {Kochan},\ and\ \citenamefont {Fabian}}]{Frank2018}%
  \BibitemOpen
  \bibfield  {author} {\bibinfo {author} {\bibfnamefont {T.}~\bibnamefont {Frank}}, \bibinfo {author} {\bibfnamefont {P.}~\bibnamefont {H\"ogl}}, \bibinfo {author} {\bibfnamefont {M.}~\bibnamefont {Gmitra}}, \bibinfo {author} {\bibfnamefont {D.}~\bibnamefont {Kochan}},\ and\ \bibinfo {author} {\bibfnamefont {J.}~\bibnamefont {Fabian}},\ }\bibfield  {title} {\bibinfo {title} {Protected pseudohelical edge states in ${\mathbb{z}}_{2}$-trivial proximitized graphene},\ }\href {https://doi.org/10.1103/PhysRevLett.120.156402} {\bibfield  {journal} {\bibinfo  {journal} {Phys. Rev. Lett.}\ }\textbf {\bibinfo {volume} {120}},\ \bibinfo {pages} {156402} (\bibinfo {year} {2018})}\BibitemShut {NoStop}%
\bibitem [{\citenamefont {Srivastava}\ \emph {et~al.}(2015)\citenamefont {Srivastava}, \citenamefont {Sidler}, \citenamefont {Allain}, \citenamefont {Lembke}, \citenamefont {Kis},\ and\ \citenamefont {Imamoğlu}}]{Srivastava2015}%
  \BibitemOpen
  \bibfield  {author} {\bibinfo {author} {\bibfnamefont {A.}~\bibnamefont {Srivastava}}, \bibinfo {author} {\bibfnamefont {M.}~\bibnamefont {Sidler}}, \bibinfo {author} {\bibfnamefont {A.~V.}\ \bibnamefont {Allain}}, \bibinfo {author} {\bibfnamefont {D.~S.}\ \bibnamefont {Lembke}}, \bibinfo {author} {\bibfnamefont {A.}~\bibnamefont {Kis}},\ and\ \bibinfo {author} {\bibfnamefont {A.}~\bibnamefont {Imamoğlu}},\ }\bibfield  {title} {\bibinfo {title} {Valley zeeman effect in elementary optical excitations of monolayer wse2},\ }\href {https://doi.org/10.1038/nphys3203} {\bibfield  {journal} {\bibinfo  {journal} {Nature Physics}\ }\textbf {\bibinfo {volume} {11}},\ \bibinfo {pages} {141–147} (\bibinfo {year} {2015})}\BibitemShut {NoStop}%
\bibitem [{\citenamefont {Tong}\ \emph {et~al.}(2016)\citenamefont {Tong}, \citenamefont {Gong}, \citenamefont {Wan},\ and\ \citenamefont {Duan}}]{Tong2016}%
  \BibitemOpen
  \bibfield  {author} {\bibinfo {author} {\bibfnamefont {W.-Y.}\ \bibnamefont {Tong}}, \bibinfo {author} {\bibfnamefont {S.-J.}\ \bibnamefont {Gong}}, \bibinfo {author} {\bibfnamefont {X.}~\bibnamefont {Wan}},\ and\ \bibinfo {author} {\bibfnamefont {C.-G.}\ \bibnamefont {Duan}},\ }\bibfield  {title} {\bibinfo {title} {Concepts of ferrovalley material and anomalous valley hall effect},\ }\bibfield  {journal} {\bibinfo  {journal} {Nature Communications}\ }\textbf {\bibinfo {volume} {7}},\ \href {https://doi.org/10.1038/ncomms13612} {10.1038/ncomms13612} (\bibinfo {year} {2016})\BibitemShut {NoStop}%
\bibitem [{\citenamefont {Lee}\ \emph {et~al.}(2025)\citenamefont {Lee}, \citenamefont {Fu}, \citenamefont {Liu}, \citenamefont {Lee},\ and\ \citenamefont {Ang}}]{Lee2025}%
  \BibitemOpen
  \bibfield  {author} {\bibinfo {author} {\bibfnamefont {K.~W.}\ \bibnamefont {Lee}}, \bibinfo {author} {\bibfnamefont {P.-H.}\ \bibnamefont {Fu}}, \bibinfo {author} {\bibfnamefont {J.-F.}\ \bibnamefont {Liu}}, \bibinfo {author} {\bibfnamefont {C.~H.}\ \bibnamefont {Lee}},\ and\ \bibinfo {author} {\bibfnamefont {Y.~S.}\ \bibnamefont {Ang}},\ }\bibfield  {title} {\bibinfo {title} {Valley gapless semiconductor: Models and applications},\ }\href {https://doi.org/10.1103/5bfl-v1qk} {\bibfield  {journal} {\bibinfo  {journal} {Phys. Rev. B}\ }\textbf {\bibinfo {volume} {111}},\ \bibinfo {pages} {235430} (\bibinfo {year} {2025})}\BibitemShut {NoStop}%
\bibitem [{\citenamefont {Mannaï}\ and\ \citenamefont {Haddad}(2020)}]{Mannai2020}%
  \BibitemOpen
  \bibfield  {author} {\bibinfo {author} {\bibfnamefont {M.}~\bibnamefont {Mannaï}}\ and\ \bibinfo {author} {\bibfnamefont {S.}~\bibnamefont {Haddad}},\ }\bibfield  {title} {\bibinfo {title} {Strain tuned topology in the haldane and the modified haldane models},\ }\href {https://doi.org/10.1088/1361-648X/ab73a1} {\bibfield  {journal} {\bibinfo  {journal} {Journal of Physics: Condensed Matter}\ }\textbf {\bibinfo {volume} {32}},\ \bibinfo {pages} {225501} (\bibinfo {year} {2020})}\BibitemShut {NoStop}%
\bibitem [{\citenamefont {Kim}\ \emph {et~al.}(2021)\citenamefont {Kim}, \citenamefont {Beach}, \citenamefont {Lee}, \citenamefont {Ono}, \citenamefont {Rasing},\ and\ \citenamefont {Yang}}]{Kim2021}%
  \BibitemOpen
  \bibfield  {author} {\bibinfo {author} {\bibfnamefont {S.~K.}\ \bibnamefont {Kim}}, \bibinfo {author} {\bibfnamefont {G.~S.~D.}\ \bibnamefont {Beach}}, \bibinfo {author} {\bibfnamefont {K.-J.}\ \bibnamefont {Lee}}, \bibinfo {author} {\bibfnamefont {T.}~\bibnamefont {Ono}}, \bibinfo {author} {\bibfnamefont {T.}~\bibnamefont {Rasing}},\ and\ \bibinfo {author} {\bibfnamefont {H.}~\bibnamefont {Yang}},\ }\bibfield  {title} {\bibinfo {title} {Ferrimagnetic spintronics},\ }\href {https://doi.org/10.1038/s41563-021-01139-4} {\bibfield  {journal} {\bibinfo  {journal} {Nature Materials}\ }\textbf {\bibinfo {volume} {21}},\ \bibinfo {pages} {24–34} (\bibinfo {year} {2021})}\BibitemShut {NoStop}%
\bibitem [{\citenamefont {Baltz}\ \emph {et~al.}(2018)\citenamefont {Baltz}, \citenamefont {Manchon}, \citenamefont {Tsoi}, \citenamefont {Moriyama}, \citenamefont {Ono},\ and\ \citenamefont {Tserkovnyak}}]{Baltz2018}%
  \BibitemOpen
  \bibfield  {author} {\bibinfo {author} {\bibfnamefont {V.}~\bibnamefont {Baltz}}, \bibinfo {author} {\bibfnamefont {A.}~\bibnamefont {Manchon}}, \bibinfo {author} {\bibfnamefont {M.}~\bibnamefont {Tsoi}}, \bibinfo {author} {\bibfnamefont {T.}~\bibnamefont {Moriyama}}, \bibinfo {author} {\bibfnamefont {T.}~\bibnamefont {Ono}},\ and\ \bibinfo {author} {\bibfnamefont {Y.}~\bibnamefont {Tserkovnyak}},\ }\bibfield  {title} {\bibinfo {title} {Antiferromagnetic spintronics},\ }\href {https://doi.org/10.1103/RevModPhys.90.015005} {\bibfield  {journal} {\bibinfo  {journal} {Rev. Mod. Phys.}\ }\textbf {\bibinfo {volume} {90}},\ \bibinfo {pages} {015005} (\bibinfo {year} {2018})}\BibitemShut {NoStop}%
\bibitem [{\citenamefont {Jungwirth}\ \emph {et~al.}(2016)\citenamefont {Jungwirth}, \citenamefont {Marti}, \citenamefont {Wadley},\ and\ \citenamefont {Wunderlich}}]{Jungwirth2016}%
  \BibitemOpen
  \bibfield  {author} {\bibinfo {author} {\bibfnamefont {T.}~\bibnamefont {Jungwirth}}, \bibinfo {author} {\bibfnamefont {X.}~\bibnamefont {Marti}}, \bibinfo {author} {\bibfnamefont {P.}~\bibnamefont {Wadley}},\ and\ \bibinfo {author} {\bibfnamefont {J.}~\bibnamefont {Wunderlich}},\ }\bibfield  {title} {\bibinfo {title} {Antiferromagnetic spintronics},\ }\href {https://doi.org/10.1038/nnano.2016.18} {\bibfield  {journal} {\bibinfo  {journal} {Nature Nanotechnology}\ }\textbf {\bibinfo {volume} {11}},\ \bibinfo {pages} {231–241} (\bibinfo {year} {2016})}\BibitemShut {NoStop}%
\bibitem [{\citenamefont {Weiße}\ \emph {et~al.}(2006)\citenamefont {Weiße}, \citenamefont {Wellein}, \citenamefont {Alvermann},\ and\ \citenamefont {Fehske}}]{Weise2006}%
  \BibitemOpen
  \bibfield  {author} {\bibinfo {author} {\bibfnamefont {A.}~\bibnamefont {Weiße}}, \bibinfo {author} {\bibfnamefont {G.}~\bibnamefont {Wellein}}, \bibinfo {author} {\bibfnamefont {A.}~\bibnamefont {Alvermann}},\ and\ \bibinfo {author} {\bibfnamefont {H.}~\bibnamefont {Fehske}},\ }\bibfield  {title} {\bibinfo {title} {The kernel polynomial method},\ }\href {https://doi.org/10.1103/RevModPhys.78.275} {\bibfield  {journal} {\bibinfo  {journal} {Reviews of Modern Physics}\ }\textbf {\bibinfo {volume} {78}},\ \bibinfo {pages} {275} (\bibinfo {year} {2006})},\ \bibinfo {note} {publisher: American Physical Society}\BibitemShut {NoStop}%
\end{thebibliography}

%


\onecolumngrid
\renewcommand{\thefigure}{S\arabic{figure}}
\renewcommand{\theequation}{S\arabic{equation}}
\renewcommand{\thetable}{S\arabic{table}}

\setcounter{figure}{0}
\setcounter{equation}{0}
\setcounter{table}{0}

\newpage
\vskip 1cm
{\centering
\large \bf Supplementary Material For ``Staggered orbital magnetization from itinerant electrons: orbital antiferro- and ferrimagnetic phases"\par}
\vskip 1cm

\section{Reciprocal-space Hamiltonian and Orbital Magnetization}  

Here, we consider the ${\bf k}$-space representation (bulk 2D system) of the Hamiltonian given in Eq. (\ref{Htot}) of the main text,
\begin{eqnarray}
\mathcal{H}({\bf k})&=&\sigma_z \Delta+\sigma_0d_0({\bf k})+\sigma_xd_x({\bf k})+\sigma_yd_y({\bf k})+\sigma_{\alpha}d_z({\bf k}),
\label{HHaldanek}
\end{eqnarray}
where, $\sigma_{0,x,y,z}$ are Pauli matrices spanned in sublattice basis $\{\ket{A}, \ket{B} \}$. We also defined the functions $d_0({\bf k})=2t_2\cos(\phi)\sum^3_{i=1}\cos({\bf k}\cdot {\bf b}_i)$, $d_x({\bf k})=t_1\sum^3_{i=1}\cos({\bf k}\cdot {\bf a}_i)$, $d_y({\bf k})=t_1\sum^3_{i=1}\sin({\bf k}\cdot {\bf a}_i)$, and $d_z({\bf k})=-2t_2\sin (\phi)\sum^3_{i=1}\sin({\bf k}\cdot {\bf b}_i)$, where ${\bf a}_1=a(0,1)$, ${\bf a}_2=a(\sqrt{3}/2,-1/2)$ and, ${\bf a}_3=a(-\sqrt{3}/2,-1/2)$ are the three NN vectors. The vectors ${\bf b}_i$ that appear in the functions $d_0({\bf k})$ and $d_3({\bf k})$ are given by ${\bf b}_1={\bf a}_2-{\bf a}_3$, ${\bf b}_2={\bf a}_3-{\bf a}_1$, and ${\bf b}_3={\bf a}_1-{\bf a}_2$. Note that the Pauli matrix appearing in the last term of Eq. (\ref{HHaldanek}) for the standard Haldane model differs from that in the modified Haldane model. In the standard Haldane model, the last term reads $\sigma_zd_z({\bf k})$, whereas in the modified Haldane model it can be cast as $\sigma_0d_z({\bf k})$. This difference follows from distinct conventions for the phases of complex hoppings $\nu^{\rm H}_{{\bf i}{\bf j}}$ and $\nu^{\rm mH}_{{\bf i}{\bf j}}$ adopted in Eq. (\ref{Htot}) of the main text. 

\begin{figure}[h]
    \centering
    \includegraphics[width=1.0\linewidth]{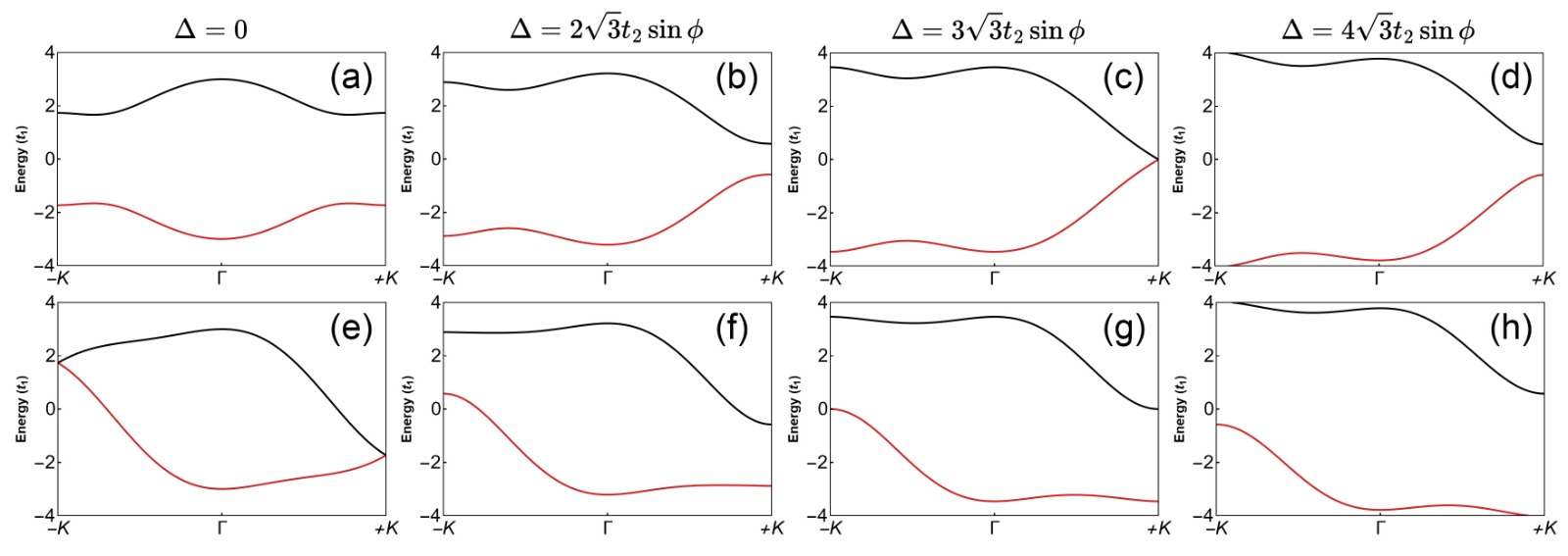}
   \caption{{\bf k}-space energy spectra for the Haldane model (a-d) and the modified Haldane model (e-h). Here we set, $t_1=1$ eV, $t_2=t_1/3$, $\phi=\pi/2$ and $\Delta=0$ (a,e), $2\sqrt{3}t_2\sin(\phi)$ (b,f), $3\sqrt{3}t_2\sin(\phi)$ (c,g), $4\sqrt{3}t_2\sin(\phi)$ (d,h).} \label{figS1}
\end{figure}

In the bottom panels of Fig. \ref{fig1}, one shows the electronic band gap of both models across the $\phi-\Delta$ parameter space. In the standard Haldane model, the complex phase in the term $t_2$ opens a topological direct gap with a quantized Chern number. Upon turning on $\Delta$, the band gap decreases and closes at the topological phase transition point. As $\Delta$ is further increased, the band gap reopens but with a trivial (zero) Chern number (see bottom panel of Fig. \ref{fig1}(a)). In the case of the modified Haldane model, the system remains gapless when the $t_2$ term is introduced and $\Delta = 0$. Indeed, the effect of the $t_2$ term on the electronic spectrum is to produce a valley-dependent energy shift of the Dirac crossings. When $\Delta$ is turned on, the Dirac cones acquire an energy splitting (local gap), but for small values of $\Delta$ the system remains metallic. For sufficiently large $\Delta$, the system becomes an insulator with an indirect gap and zero Chern number (see bottom panel of Fig. \ref{fig1}(b)). In Fig. \ref{figS1}, we illustrate the evolution of the electronic band structures for both models. 

Within the framework of the so-called modern theory of orbital magnetization \cite{Xiao2005, Thonhauser2005}, one can cast it in terms of the periodic part of the Bloch states $\ket{u_{n{\bf k}}}$ with energy $\epsilon_{n{\bf k}}$: $M_z=- \frac{e}{2\hbar c} {\rm Im} \sum_{n}\int_{\epsilon_{n{\bf k}}\le E_F}\frac{d{\bf k}}{(2\pi)^2} \bra{\boldsymbol{\nabla}_{\bf k}u_{n{\bf k}}}\boldsymbol{\times}\left(\hat{H}_{\bf k}+\mathbb{1}(\epsilon_{n{\bf k}}-2E_{\rm F})\right) \ket{\boldsymbol{\nabla}_{\bf k}u_{n{\bf k}}}$. This formula can be used to benchmark the real-space implementation of the total orbital magnetization $M_A+M_B$ for sufficiently large systems, as done in Ref. \cite{Vidarte2026}.

\section{Low energy theory and symmetry analysis}

One can also expand the Hamiltonian of Eq. (\ref{HHaldanek}) around the valleys of the Brillouin zone to obtain the following Dirac Hamiltonian,
\begin{eqnarray}
\mathcal{H}({\bf q})=\hbar v_{\rm F}(\tau_zq_x\sigma_x+q_y\sigma_y)+\Delta\sigma_z+m\sigma_{\alpha}\tau_z,
\label{HDirac}
\end{eqnarray}
where, $v_{\rm F}=3at_1/(2\hbar)$ is the Fermi velocity and $m=3\sqrt{3}t_2\sin(\phi)$ is the mass term at valleys associated with complex NNN hoppings. ${\bf q}$ is wave-vector relative to the valleys, and $\tau_z=+1(-1)$ for the ${\bf K}$ (${\bf K'}$) valley. Again $\sigma_{\rm H}=\sigma_z$ for standard Haldane ($\alpha=\rm{H}$) model, and $\sigma_{\rm mH}=\sigma_0$ for the modified Haldane ($\alpha=\rm{mH}$) model. 

The two models exhibit distinct symmetry properties. For simplicity, we consider the Hamiltonian at the ${\bf K}$ and ${\bf K'}$ points, where the Dirac term vanishes ${\bf q}=0$. Spatial inversion acts on the sublattice and valley degrees of freedom as $P: \ket{A(B)}\rightarrow \ket{B(A)}$, and $P: \ket{{\bf K}({\bf K'})}\rightarrow \ket{{\bf K'}({\bf K})}$. The time-reversal operator acts as $T: \ket{A(B)}\rightarrow \ket{A(B)}$, and $T: \ket{{\bf K}({\bf K'})}\rightarrow \ket{{\bf K'}({\bf K})}$. We then obtain their action on the Pauli matrices associated with the sublattice and valley degrees of freedom. In particular,
\begin{eqnarray}
P:
\left\{
\begin{aligned}
\sigma_z &\rightarrow -\sigma_z, \\
\tau_z &\rightarrow -\tau_z.
\end{aligned}
\right.
\end{eqnarray}

and 

\begin{eqnarray}
T:
\left\{
\begin{aligned}
\sigma_z &\rightarrow \sigma_z, \\
\tau_z &\rightarrow -\tau_z.
\end{aligned}
\right.
\end{eqnarray}
From these relations, one can easily construct the symmetry analysis presented in the main text for the mass term in standart Haldane model, modified Haldane model and sublattice potential.

\section{Spectral method}

Following the modern formulation of orbital magnetization in finite system \cite{Bianco2013, Vidarte2026}, the total 
magnetization of a system possessing two inequivalent sublattices, $\sigma=A$ and $B$, can be partitioned into sublattice-resolved contributions as
\begin{equation}
M_{\sigma} = -\frac{ie}{2\hbar c\,S}
\sum_{\epsilon_n<E_F}
\langle \varphi_n |\, \hat{\mathbf r}_{\sigma}\times[\hat {\mathcal{H}},\hat{\mathbf r}_{\sigma}] \,| \varphi_n \rangle ,
\label{eq:s1}
\end{equation}
where $\hat{\mathbf r}_{\sigma}$ is the position operator of sites from sublattice $\sigma=A(B)$, $S$ is the sample area, $\hat{\mathcal{H}}$ is the single-particle Hamiltonian, $\epsilon_n$ and $|\varphi_n\rangle$ are the eigenvalues and eigenvectors of $\hat {\mathcal{H}}$, respectively, and $E_F$ is the Fermi energy. 
We define operators \cite{Vidarte2026}
\begin{eqnarray}
\hat{\mathscr{M}}_{\sigma}=-\frac{ie}{2\hbar c S}(\hat{x}_{\sigma}\hat {\mathcal{H}}\hat{y}_{\sigma}-\hat{y}_{\sigma}\hat{\mathcal{H}}\hat{x}_{\sigma})\,\, ,
\label{eq:s2}
\end{eqnarray}
and Eq.(\ref{eq:s1}) can be written as as $M_\sigma=\sum_{\epsilon_n<E_F}\langle\varphi_n|\hat{\mathscr{M}}_{\sigma}|\varphi_n\rangle$. Within the framework of the Chebyshev polynomial expansion \cite{Weise2006}, the sublattice-resolved \emph{magnetization spectral density} is defined as \cite{Vidarte2026}:
\begin{equation}
m_{\sigma}(E)= \mathrm{Tr}[\hat{\mathscr{M}}_{\sigma}\,\delta(E-\hat{\mathcal{H}})].
\label{eq:4}
\end{equation}
Knowledge of this quantity allows to retrieve the sublattice orbital magnetization via a straightforward and quick integration according to $M_{\sigma}(E_F)=\int^{E_F} m_{\sigma}(E)\,dE$.  
To efficiently evaluate $m_{\sigma}(E)$, we expand $\delta(E-\hat{\mathcal{H}})$ in Chebyshev polynomials after
rescaling $\hat {\mathcal{H}}\!\to\!\tilde{\mathcal{H}}$ and $E\!\to\!\tilde{E}\in[-1,1]$.
The expansion coefficients are the moments $\mu^{\sigma}_n=\mathrm{Tr}[\hat{\mathscr{M}}_{\sigma} T_n(\tilde{\mathcal{H}})]$, computed iteratively via the Chebyshev recursion procedure. 
To further speed up the Chebyshev scheme, the traces are computed stochastically according to
$\mu^{\sigma}_n\!\approx\!R^{-1}\sum_{r=1}^R\langle r|\hat{\mathscr{M}}_{\sigma} T_n(\tilde{\mathcal{H}})|r\rangle$, where $|r\rangle$ is the $r$-th realization of a random vector  \cite{Weise2006}. 
The resulting expression for $m_{\sigma}(E)$ scales linearly with the number of random vectors $R$, the number of lattice sites, $N$, and the number of Chebyshev iterations, $M$.

All numerical results obtained via the spectral method are based on the following parameters: $N \approx 10^6$ sites, $M = 1024$, and $R = 5000$.

\section{Details on the low-energy staggered orbital magnetization in the trivial insulating phase of the standard Haldane model}

Here, we provide further details on the low-energy formulation of the bulk standard Haldane model in {\bf k}-space to qualitatively explain the orbital antiferromagnetic phase exhibited by the trivial insulator for $\phi=\pi/2$, as shown in Fig. 4(a) of the main text. The matrix representation in the basis $\left\{\ket{A}, \ket{B} \right\}$ is given by

\begin{eqnarray}
\mathcal{H}_{\rm H}({\bf q})=\hbar v_{\rm F}(\tau_zq_x\sigma_x+q_y\sigma_y)+\Delta\sigma_z+m\sigma_z\tau_z=\begin{bmatrix}
   \Delta-\tau m & \hbar v_{\rm F}\left(\tau q_x-iq_y\right) \\
   \hbar v_{\rm F}\left(\tau q_x+iq_y\right) & -(\Delta-\tau m) 
\end{bmatrix}.
\label{Hmat}
\end{eqnarray}
The energy spectrum of the Hamiltonian is given by $\tilde{\epsilon}_{c(v){\bf q}}=\pm \tilde{\varepsilon}_q$, where $\tilde{\varepsilon}_q=\sqrt{v^2_{\rm F}\hbar^2q^2+\left(\Delta-\tau m\right)^2}$. It is possible to write the projector for the valence band and for the conduction band,
\begin{eqnarray}
\mathcal{P}_{v(c)}=\frac{1}{2}\left(\mathbb{1}\mp\frac{\mathcal{H}_{\rm H}({\bf q})}{\tilde{\varepsilon}_q}\right), 
\label{Pvc}
\end{eqnarray}
and the sublattice projectors,
\begin{eqnarray}
\mathcal{S}_{A(B)}=\frac{\mathbb{1}\pm \sigma_z}{2}. 
\label{SAB}
\end{eqnarray}
Therefore, the probability weight associated with each band projected on the sublattice can be expressed as $w_{n\sigma{\bf q}}=\text{Tr}\left[\mathcal{S}_{\sigma} \mathcal{P}_{n}\right]$. It is also convenient to define the quantity, 
\begin{eqnarray}
\Lambda_{\bf q}\equiv \text{Im}\left[ \braket{\partial_{x}u_{v{\bf q}}}{u_{c{\bf q}}} \braket{u_{c{\bf q}}}{\partial_{y}u_{v{\bf q}}}\right]. 
\end{eqnarray}
In the two-band model, it is directly related to the Berry curvature of the valence band,
\begin{eqnarray}
\tilde{\Omega}_{v{\bf q}}=-2\text{Im}\left[\braket{\partial_x u_{v{\bf q}}}{\partial_y u_{v{\bf q}}}\right]=-2\Lambda_{\bf q},
\end{eqnarray}
where we use the resolution of the identity $\mathbb{1}=\ket{u_{c{\bf q}}}\bra{u_{c{\bf q}}}+\ket{u_{v{\bf q}}}\bra{u_{v{\bf q}}}$ in the last equality.

The {\bf k}-space formula for the modern theory of orbital magnetization reads, $M_z=- \frac{e}{2\hbar c} {\rm Im} \sum_{n}\int_{\epsilon_{n{\bf k}}\le E_F}\frac{d{\bf k}}{(2\pi)^2} \bra{\boldsymbol{\nabla}_{\bf k}u_{n{\bf k}}}\boldsymbol{\times}\left(\hat{H}_{\bf k}+\mathbb{1}(\epsilon_{n{\bf k}}-2E_{\rm F})\right) \ket{\boldsymbol{\nabla}_{\bf k}u_{n{\bf k}}}$. For a single-occupied band in a two-band model, one may use the projectors from Eqs. (\ref{Pvc}, \ref{SAB}) to write the sublattice-projected orbital magnetization as
\begin{eqnarray}
M_{\sigma}=\frac{e}{\hbar c}\sum_{\tau=\pm 1} \int\frac{d{\bf q}}{(2\pi)^2}\mathcal{I}_{\sigma{\bf q}}, \label{Mklattice}
\end{eqnarray}
where, 
\begin{eqnarray}
\mathcal{I}_{\sigma{\bf q}}=\left[\tilde{\epsilon}_{c{\bf q}}w_{v\sigma{\bf q}}+(\tilde{\epsilon}_{v{\bf q}}-2E_{\rm F})w_{c\sigma{\bf q}} \right]\Lambda_{\bf q}. \label{Iklattice}
\end{eqnarray}

One can obtain the explicit expressions for the probability weights in the standard Haldane model by using Eqs. (\ref{Pvc}) and (\ref{SAB}): $w_{v(c)A{\bf q}}=(1/2)\left[1\mp(\Delta-\tau m)/\tilde{\varepsilon}_{q}\right]$ and $w_{v(c)B{\bf q}}=(1/2)\left[1\pm(\Delta-\tau m)/\tilde{\varepsilon}_{q}\right]$. Using these expressions, Eq. (\ref{Iklattice}) can be cast as:
\begin{eqnarray}
\mathcal{I}_{A(B){\bf q}}=
\left[\mp\left(\Delta-\tau m\right) - E_{\rm F}\left(1\pm \frac{(\Delta-\tau m)}{\tilde{\varepsilon}_q} \right)\right]\Lambda_{\bf q}
=\pm\frac{1}{2}\left[\left(\Delta-\tau m\right) \pm E_{\rm F}\left(1\pm \frac{(\Delta-\tau m)}{\tilde{\varepsilon}_q} \right)\right]\tilde{\Omega}_{v{\bf q}}. \label{Iab}
\end{eqnarray}
The valence-band Berry curvature can be straightforwardly computed from the eigenvectors of Eq. (\ref{Hmat}):
\begin{eqnarray}
\tilde{\Omega}_{v{\bf q}}=\tau\hbar^2v^2_{\rm F}\frac{(\Delta-\tau m)}{2\tilde{\varepsilon}^3_q}, \label{omegav}
\end{eqnarray}
Thus, substituting Eq. (\ref{omegav}) into Eqs. (\ref{Iab}, \ref{Mklattice}) yields the sublattice-resolved orbital magnetization of the trivial insulating phase of the Haldane model. 

To calculate the total and staggered orbital magnetizations, we define the following quantities:
\begin{eqnarray}
\mathcal{I}_{z{\bf q}}&\equiv& \mathcal{I}_{\rm A {\bf q}}+\mathcal{I}_{\rm B {\bf q}} \nonumber \\
&=&E_{\rm F}\tilde{\Omega}_{v{\bf q}} \label{Iz}
\end{eqnarray}
and
\begin{eqnarray}
\mathcal{I}^s_{z{\bf q}}&\equiv& \mathcal{I}_{\rm A{\bf q}}-\mathcal{I}_{\rm B{\bf q}} \nonumber \\
&=&(\Delta-\tau m)\left(1+\frac{E_{\rm F}}{\tilde{\varepsilon}_q}\right)\tilde{\Omega}_{v{\bf q}}. \label{Isz}
\end{eqnarray}
With these definitions, one may write: 
\begin{eqnarray}
M_z=\frac{e}{\hbar c}\sum_{\tau=\pm 1} \int\frac{d{\bf q}}{(2\pi)^2}\left[\mathcal{I}_{A{\bf q}}+\mathcal{I}_{B{\bf q}}\right], \label{Mzk}
\end{eqnarray}
and 
\begin{eqnarray}
M^s_z=\frac{e}{\hbar c}\sum_{\tau=\pm 1} \int\frac{d{\bf q}}{(2\pi)^2}\left[\mathcal{I}_{A{\bf q}}-\mathcal{I}_{B{\bf q}}\right]. \label{Mszk}
\end{eqnarray}

\subsection*{Orbital Magnetization}

In the case of uniform orbital magnetization, one obtains the expected result for the standard Haldane model in the trivial insulating phase for $\phi=\pi/2$:
\begin{eqnarray}
M_z=\frac{E_{\rm F}e}{\hbar c}2\sum_{\tau=\pm 1} \int\frac{d{\bf q}}{(2\pi)^2}\tilde{\Omega}_{v{\bf q}} \propto E_{\rm F}C=0. \label{MzResult}
\end{eqnarray}
Note that the finite orbital magnetization that appears inside the trivial insulating phase for $\phi\neq\pi/2$ (see Section \ref{SMV}) is not captured by the low-energy calculation, suggesting that points in the Brillouin zone away from the valleys may also be relevant in this case. 

\subsection*{Staggered Orbital Magnetization}

The contribution from the term proportional to the Fermi energy in Eq. (\ref{Isz}) cancels when summed over the valleys,
\begin{eqnarray}
\frac{e}{\hbar c}\sum_{\tau=\pm 1} \int\frac{d{\bf q}}{(2\pi)^2}E_{\rm F}\frac{(\Delta-\tau m)}{\tilde{\varepsilon}_q}\tilde{\Omega}_{v{\bf q}}=0.
\end{eqnarray}
Only the term independent of the Fermi energy contributes to the staggered orbital magnetization. By extrapolating the short-wavelength cutoff to infinity in the momentum integral one obtains the following result:
\begin{eqnarray}
M^s_z=\frac{e}{\hbar c}\sum_{\tau=\pm 1} \int\frac{d{\bf q}}{(2\pi)^2}(\Delta-\tau m)\tilde{\Omega}_{v{\bf q}}=-\frac{e}{\hbar c}\frac{(\Delta- m)-(\Delta+ m)}{4\pi}=\frac{e}{\hbar c}\frac{m}{2\pi}. \label{MszResult}
\end{eqnarray}.

Then, the results in Eqs. (\ref{MzResult}) and (\ref{MszResult}) show that the orbital antiferromagnetic insulating state of the standard Haldane model in the trivial phase for $\phi=\pi/2$ can be explained within the framework of low-energy theory.

\section{Additional results on orbital magnetization and the density of states \label{SMV}}
\begin{figure}[h]
    \centering
    \includegraphics[width=0.55\linewidth]{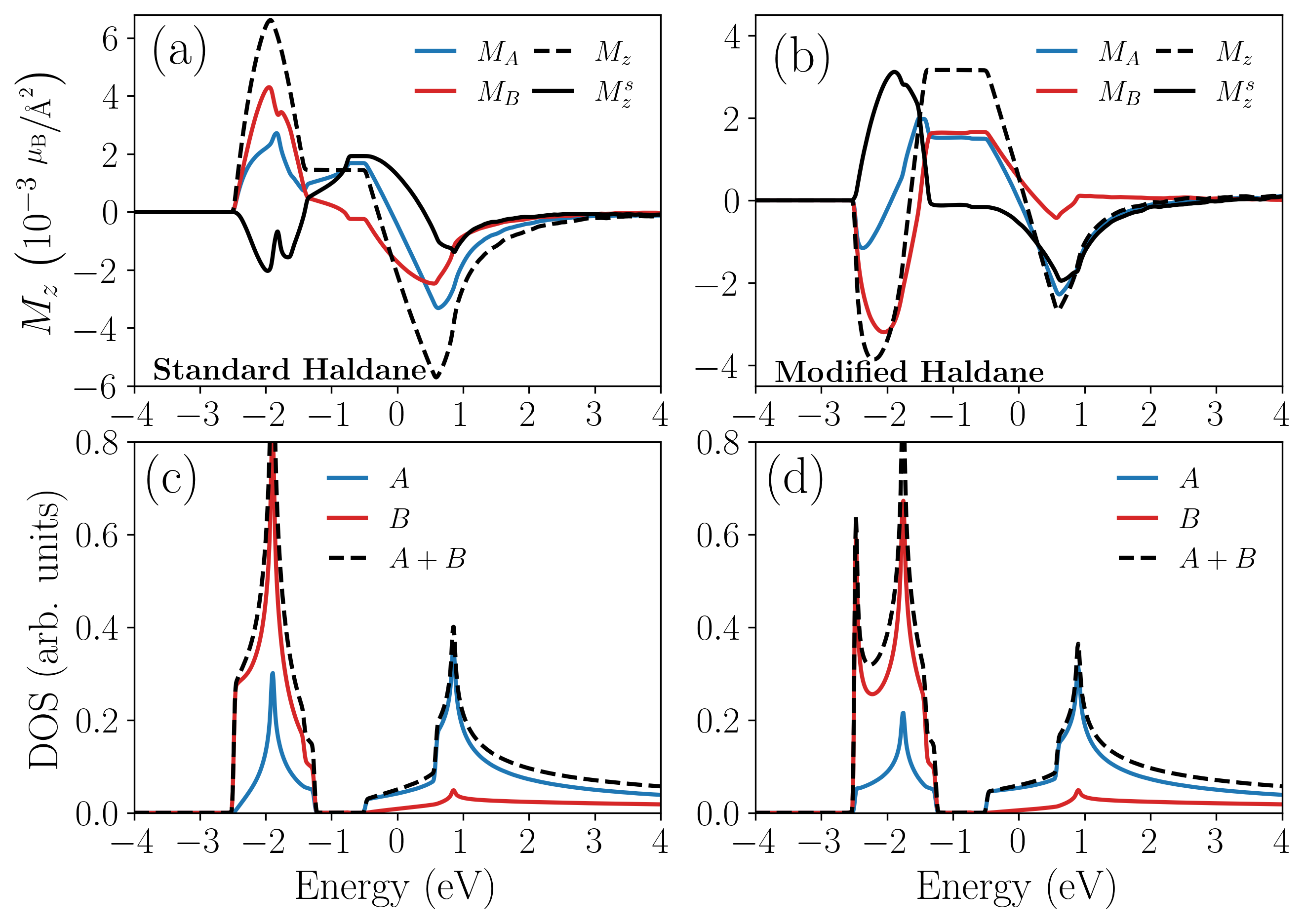}
   \caption{Sublattice-resolved ($M_A$ and $M_B$), total ($M_z$), and staggered ($M_z^s$) orbital magnetizations for the standard and modified Haldane models in panels (a) and (b), respectively, and the corresponding density of states in panels (c) and (d). Here, we set $t_1=1 \text{eV}$, $t_2=t_1/3$, $\phi=0.1\pi$, and $\Delta=3t_2$. 
   } \label{figS2}
\end{figure}

In Figure \ref{figS2}, we show the sublattice-resolved, staggered, and total orbital magnetizations for $\Delta>3\sqrt{3}\sin(\phi)t_2$ in the absence of particle-hole symmetry in the DOS. Within the insulating band gap of both models, $M_{\rm A}\neq M_{\rm B}$; consequently, both $M_z$ and $M^s_z$ are finite, characterizing an itinerant orbital ferromagnetic insulating phase. Note that in the modified Haldane model for the Fermi energy inside the insulating region, the staggered component is finite, although very small. 

\end{document}